\RequirePackage{fix-cm}
\documentclass[11pt,a4paper]{article}
\usepackage{color}
\usepackage{amsmath}
\usepackage{multirow}
\usepackage{amssymb}
\usepackage{graphicx}
\usepackage{float}
\usepackage[a4paper]{geometry}
\usepackage{subfigure}
\usepackage{multicol}
\usepackage{epstopdf}
\usepackage{authblk}

\newcommand{\dd}{\mathrm{d}}
\def\WW{{\mathbb{W}}}
\def\MM{{\mathbb{M}}}
\def\dd{{\rm d}}
\def\half{\mbox{$1\over2$}}

\def\thir{\mbox{$1\over3$}}
\def\quar{\mbox{$1\over4$}}
\def\3quar{\mbox{$3\over4$}}
\def\5quar{\mbox{$5\over4$}}
\def\7quar{\mbox{$7\over4$}}

\newcommand{\be}{\begin{equation}}
\newcommand{\ee}{\end{equation}}
\def\half{{\mbox{$1\over2$}}}

\newcommand{\pd}[2]{\frac{\partial #1}{\partial #2}}

\def\hze{{\hat\zeta}}
\def\hM{{\hat M}}

\def\cL{{\cal L}}
\def\cI{{\cal I}}
\def\cP{{\cal P}}
\def\hcL{{\hat{\cal L}}}
\def\hcP{{\hat{\cal P}}}

\title{Dynamics of levitation during
rolling over a thin viscous film}

\author[1]{S. Chen\thanks{Corresponding author. Email: \texttt{jack\_chen03@sjtu.edu.cn}}}
\author[2]{C. Liu}
\author[3]{Neil J. Balmforth}
\author[2]{S. Green}
\author[2]{B. Stoeber}

\affil[1]{%
  Zhiyuan College \\
  Shanghai Jiao Tong University \\
  Shanghai 200240, China \\
}

\affil[2]{%
  Department of Mechanical Engineering \\
  University of British Columbia \\
  Vancouver, BC V6T 1Z4, Canada
}

\affil[3]{%
  Department of Mathematics \\
  University of British Columbia \\
  Vancouver, BC V6T 1Z2, Canada
}

\date{}

\begin{document}
\maketitle

\begin{abstract} 
A mathematical model is derived for the dynamics of a cylinder,
or wheel, rolling over a thin viscous
film. The model combines the Reynolds lubrication equation for the fluid
with an equation of motion for the wheel.
Two asymptotic limits are studied in detail
to interrogate the dynamics of levitation:
an infinitely wide wheel and a relatively narrow one. In both cases
the front and back of the fluid-filled gap are either straight or nearly
so. To bridge the gap between these two asymptotic limits,
wheels of finite width are considered, introducing a further
simplying approximation: although
the front and back are no longer expected to remain straight
for a finite width, the
footprint of the fluid-filled gap is still taken to be rectangular,
with boundary conditions imposed at the
front and back in a wheel-averaged sense.
The Reynolds equation can then be solved by separation of variables.
For wider wheels, with a large amount of incoming flux
or a relatively heavy loading of the wheel,
the system is prone to flooding by back flow with fluid unable to
pass underneath. Otherwise steady planing states are achieved.
Both lift-off and touch-down are explored for a wheel rolling over
a film of finite length. Theoretical predictions are compared
with a set of experimental data.

\end{abstract}

\section{Introduction}

The hydrodynamic
levitation of solid objects by the flow of a viscous liquid features in a
number of classical problems, ranging from hydroplaning tires and skipping stones
to the air hockey table and fluidized beds
\cite{moore67,browne72,rosellini05,hewitt11,hinch94,weidman,davidson77}.
Somewhat similarly,
cylinders or spheres rolling over a solid surface can lift off
and become levitated
on encountering a thin viscous film coating that surface
\cite{bico09,schade11,marshall14,rollpool}.
In fact, even when held against a vertical moving belt,
rotating cylinders and spheres can remain levitated by the viscous coating
\cite{Eggers2013,ockendon24,Dalwadi2021}. 

\begin{figure}
\begin{center}
\includegraphics[width=.75\linewidth]{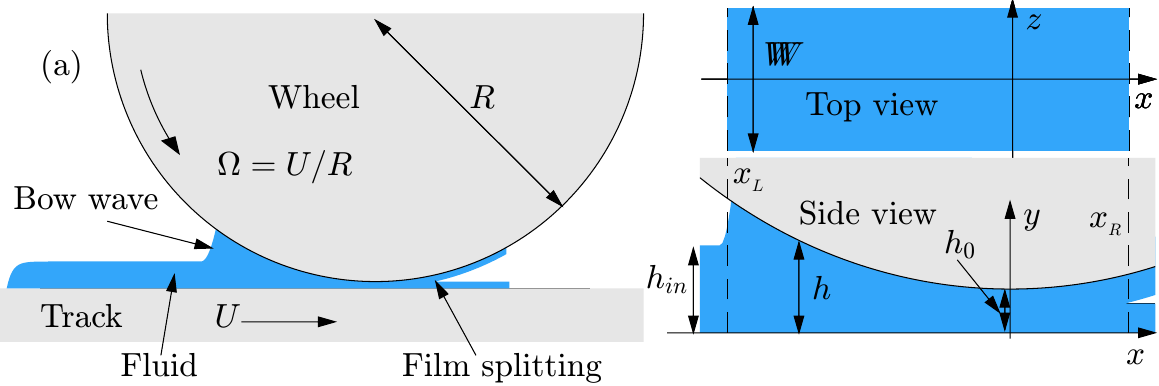}
\includegraphics[width=.95\linewidth]{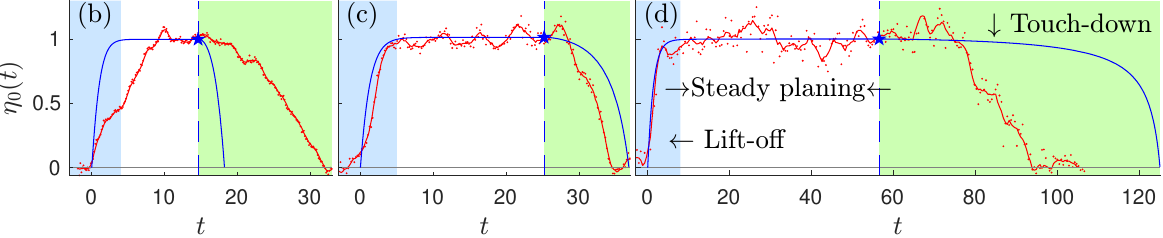}
\end{center}
\caption{\label{sketch}
  (a) A sketch of the model geometry. (b,c,d) The three phases of
  evolution arising as the wheel rolls over a film of finite length:
  lift-off, steady planing and touch-down. Displayed are
  times series of the minimum gap, $h_0(t)$, scaled by its value
  during steady planing, $h_*$ (so that $\eta_0(t)=h_0/h_*$),
  with time made dimensionless using the timescale $U/\sqrt{Rh_*}$,
  where $U$ is the rolling speed and $R$ the wheel radius.
  The vertical dashed lines and stars
  indicate the time at which the wheel reaches the end of the film.
  Three examples are shown. For each, the red dots show experimental measurements
  using the laboratory set-up from \cite{rollpool}; the red solid lines
  show a running average over 6 data points. The blue solid lines
  show corresponding predictions of the theoretical model.
  The parameter values are:
  (b) $(U,h_*,\WW)=(3$m/s$,0.012$mm$,2$cm$)$,
  (c) $(U,h_*,\WW)=(0.5$m/s$,0.059$mm$,1$cm$)$,
  (d) $(U,h_*,\WW)=(0.5$m/s$,0.176$mm$,0.5$cm$)$; in each case,
  the incoming film depth and length are $(h_{in},L_p)=(0.5,60)$mm and $R=9.55$cm.
  The dimensionless loads imposed experimentally or predicted by the model
  (as defined in \S\ref{inbetween}) are:
  (b) $\cL_0=0.022$ {\it vs.} $0.042$,
  (c) $\cL_0=0.076$ {\it vs.} $0.197$,
  (d) $\cL_0=0.129$ {\it vs.} $0.352$.
  Note that the kinematic viscosity $\nu$ of the fluid for tests (b,c) was about
  $10^{-2}$m$^2$/s; that for (d) was about $7\times10^{-4}$m$^2$/s. The Reynolds
  number based on the gap, $Uh_*/\nu$, is therefore of order
  $0.1$ or smaller.
}
\end{figure}

In the present paper and as sketched in figure \ref{sketch},
we consider a cylinder rolling over a horizontal
track, motivated by an application in engineering:
the addition of lubricants to train tracks to reduce
wheel and rail wear, noise and fuel consumption in the
rail transport industry \cite{Harmon2016,STOCK2016225}.
In this context, a pool of viscous liquid is
deposited ahead of an approaching train; the resulting interaction
with the rolling wheel, and the ensuing ``carry down'' of the lubricant,
both contribute to the desired lubrication process.

An experimental analogue of this process was conducted by
Rahmani {\it et al.} \cite{rollpool}, who showed that it was possible
for the cylinder, or wheel, to become levitated
by the viscous film even for relatively slow rolling speeds
and large loads. To rationalize this observation, they proposed 
a complementary theoretical analysis based on
Reynolds lubrication theory (film thicknesses remaining sufficiently
small for Stokes flow to apply approximately).
In that analysis, the unbounded increase of the lubrication pressure
as the gap closes between the wheel and track provides the
means for levitation under high loads and low speeds.

However, in the lubrication analsis of \cite{rollpool},
two significant simplifications were adopted
in order to arrive at a relatively simple model. First,
only time-independent, steady planing solutions were constructed.
The model did not therefore account for the time-dependent initial
lift-off and final touch-down of the wheel as it traversed a pool
of finite length (see figure \ref{sketch}(b-d)). 
Second, a crude wheel-averaged approximation
was introduced to describe the sideways flow of fluid
underneath the wheel. This latter approximation,
similar to averaging methods used in other thin-film flows
(the von Karman-Pohlhausen method), avoids the need to solve the
full two-dimensional Reynolds equation over the fluid-filled gap.
Instead, one must only find the local
wheel-average pressure along the track. 
The averaging, however, demanded the inclusion of a
free parameter in order to relate the side flux at the edges of the
wheel with the wheel-averaged pressure (see \cite{rollpool}).
The fidelity of this approximation was not tested, and instead
the free parameter was fitted, apparently successfully, using the experiments.
In the present article, we attempt to relax both approximations and
explore the dynamics of lift-off, touch-down and side flux.

%In rail transport, in order to reduce wheel and rail wear, noise and fuel consumption, a pool of viscous liquid is deposited on the track ahead of the approaching train (Harmon \& Lewis \cite{Harmon2016}; Stock \textit{et al.} \cite{STOCK2016225}; Rahmani \& Green \cite{Hatef2017}). In the process of liquid-wheel contact, the wheel is slightly lifted off by the fluid under appropriate conditions. Similar dynamics is studied in the context of a solid object levitating by a vertical moving belt coated with a thin film of viscous fluid(Eggers, Kerswell \& Mullin \cite{Eggers2013}; Mullin, Ockendon \& Ockendon \cite{Mullin2020}; Dalwadi \textit{et al.} \cite{Dalwadi2021}). 

Key difficulties in this endeavour stem from the geometry of the
fluid filled gap and the boundary conditions that must be applied at the front,
back and sides. Similar issues arise in related lubrication problems,
such as journal bearings of finite length
\cite{VIGNOLO2011,SFYRIS2012,PFEIL2023}. In the current problem,
as the wheel rolls over the pool, the fluid is partially adhered to the
front of the wheel and the film must therefore split
at a downstream meniscus, as in some other coating problems ({\it e.g.}
\cite{coyle,Decre1995,coyle97,Weinstein2004,Becerra2007});
see figure \ref{sketch}(a).
In general, the relatively short lengthscale characterizing the splitting
implies that lubrication theory cannot remain valid to describe this
region. Worse, the low pressures that can occur here often trigger
instability, cavitation or the entrainment of air
\cite{pearson60,taylor63,dowson,coyle97,Braun2010}, complicating the dynamics over
the splitting region yet further.
Similarly, where the fluid-filled gap meets the incoming viscous pool,
a relatively short bow wave must form ({\it cf.} figure \ref{sketch}(a)),
again invalidating lubrication theory.
Last, at the sides of the wheel, fluid leaves the gap to generate
the net side flux, but the sudden expansion of the fluid filled region
is potentially problematic in thin-film theory.

To rescue the analysis, all this complication is 
replaced by effective boundary conditions at the back, front
and sides of the fluid-filled gap, or lubrication zone
(see \cite{freeboundary1963,dowson,Braun2010,Taroni2012}),
an avenue we also proceed down here. Even then, however, the
geometry is not known {\it a priori}, but must be found as part of the
solution of a free-boundary problem.
Here, we take advantage of the fact that in two special limits
of the problem, some of the
complications are eased: the levitation of infinitely
wide, or relatively narrow wheels. To bridge between these two limits,
we adopt a cruder approach to deal with the geometry of the lubrication
zone: we assume that the fluid-filled gap remains roughly
rectangular, and then solve the Reynolds equation using separation of variables,
satisfying boundary conditions at the front and back
in a wheel-averaged sense.

%% these look to be about elastohydrodynamic lubrication? (not time-dep)
%There are also efforts in studying the time-dependent lubricated rolling problem using numerical methods (Durany \& García \& Vázquez \cite{DURANY1996}; Boman \& Ponthot \cite{BOMAN2002} \cite{BOMAN2004}).

%More recently, Rahmani \textit{et al.} \cite{rollpool} conducted experiments on the rolling of a cylinder over a pool of a viscous fluid, in which a lubrication theory based on the simplified Reynolds equation with an experimentally calibrated parameter is applied to approximate the pressure of the fluid underneath the cylinder.

The paper is structured as follows: In \S \ref{model}, we mathematically formulate
the lubrication theory, the equation of motion of the wheel and the boundary
conditions applying along the front, back and sides of the fluid-filled gap.
We then explore two important limits in which we may reliably simplify the
geometry of the fluid-filled region:
an infinitely wide wheel (\S \ref{wide}) and a relatively narrow one
(\S \ref{narrow}). In both cases, we build steady planing solutions
and study how the wheel lifts off towards, or touches back down from
these states.
We bridge between these two limits and consider
wheels of finite width in \S \ref{inbetween}.
Here, we also compare model predictions 
with data extracted from experiments using the set-up of \cite{rollpool}.
%A WKB approximation is applied to help locate the eigenvalues of the Sturm-Liouville problem induced by the separation of variables. To feature the dynamics, MATLAB ode15s is applied to simulate the time-dependent problem.
Finally, in \S \ref{sec:disc}, we summarize our results.
In the Appendix, we gauge the fidelity of the approximations taken in
\S \ref{inbetween}, as well as the wheel-averaged model of \cite{rollpool}.

\section{Theoretical model}\label{model}

\subsection{Lubrication theory}

We use a Cartesian coordinate to describe the problem geometry,
as sketched in figure \ref{sketch}.
The $x-$axis points in the direction of motion (axial direction),
the $y-$axis is perpendicular to the rail, and the $z-$axis
corresponds to the lateral direction. Over the nip region,
the gap thickness, 
\be
h(x,t) = R + h_0(t) - \sqrt{R^2 - x^2} \approx h_0(t) + \frac{x^2}{2R}
,
\ee
is significantly smaller than the wheel width $\WW$
and the lengthscale $\sqrt{Rh_0}$ characterizing axial variations. Hence,
following the usual reductions of lubrication theory, the leading-order
balances in the mass and momentum equations over the nip region are
\begin{align}
  u_x + v_y + w_z &= 0,   \label{1.1} \\
  p_x &= (\mu u_y)_y, \label{1.2}\\
  p_y &= 0, \label{1.3}\\
  p_z &= (\mu w_y)_y, \label{1.4}
\end{align}
where the fluid velocity and pressure are $(u,v,w)$ and $p$, respectively,
and we have used subscripts as a short-hand notation for
partial derivatives. The fluid viscosity 
is $\mu$.
The balances omit gravity and inertia, in view of the relatively high
lubrication pressure and narrowness of the gap \cite{rollpool}.

Imposing no slip on the wheel and rail, we have
\begin{align}
u&=U \ {\rm and} \ v=w=0 \ {\rm at} \ y=0, \label{1.5} \\
u&=U, \ v=h_t+Uh_x \ {\rm and} \ w=0 \ {\rm at} \ y=h(x,t) . \label{1.6}
\end{align}

The uniformity of the pressure across the gap implied by (\ref{1.3}),
indicates that (\ref{1.2}) and (\ref{1.4}) can be integrated
to find the velocity profile along the gap:
\be
\left(\begin{array}{c} u \\ w \end{array}\right)
=
\left(\begin{array}{c} U \\ 0 \end{array}\right) -
\frac{y(h-y)}{2\mu}
\left(\begin{array}{c} p_x \\ p_z \end{array}\right)
\label{1.8}
\ee
Net mass conservation across the gap indicates that
\be
h_t +
\pd{}{x}
\int_0^h  u \;\dd y +
\pd{}{z}
\int_0^h  w \;\dd y
= 0.
\ee
Introducing (\ref{1.8}) now gives the Reynolds lubrication equation,
\be
\left( x h_{0t} + Uh - \frac{h^3p_x}{12\mu} \right)_x
-
\left(\frac{h^3p_z}{12\mu} \right)_z
= 0.
\ee

\subsection{Equation of motion of the wheel}

In the experiment, the wheel is held on an axis that rotates
at the rate required for the rotation speed at the rim to
match the velocity of the track. Moreover, the track is vertical, and
the wheel axis shifts horizontally, being pushed laterally onto
the track by a pre-set load $L$. This is a little different
from a wheel that is acted upon by gravity and is
driven by a torque and reaches the track speed
due to the friction at the dry solid contact arising before the
wheel reaches the viscous pool. To model the experiment, we therefore
imagine that the wheel has an effective mass $\MM=\pi \rho_w R^2 \WW$
and the lateral
motion is determined by the simple equation of motion,
\be
\MM \ddot{h_0} = \int_{-\WW/2}^{\WW/2} \int_{x_L}^{x_R}  p(x,z,t)\; \dd x \dd z
- L \WW,
\ee
where $L$ is the load per unit width. The estimate of the effective
mass ignores the contributions from the piston connected
to the air cylinder that provides the thrust, and the bearing system that
holds the wheel in a fixed vertical position whilst allowing
horizontal motion. The parameter $\MM$ is therefore
difficult to gauge in the experiments. Instead, we treat this
quantity as a free parameter, and chiefly constrain its value
given the apparent lack of inertial effects in the experiments.

\subsection{Lubrication zone and boundary conditions}

As sketched in figure \ref{sketch}, we assume that
the lubrication zone fills the region,
$x_L<x<x_R$ and $-\frac12 \WW<z<\frac 12\WW$.
At the sides of the wheel, fluid is allowed to leave the gap,
thereby returning to atmosphere pressure (chosen to vanish
by a suitable choice of gauge). Provided there is no
back pressure, we therefore impose
\be
p(x,\pm \half \WW, t)=0
.
\label{bcsides0}
\ee

The conditions at the front and back $x=x_{L,R}$ are a little more
awkward. Lubrication theory cannot capture the relatively short horizontal
scales that develop where the incoming pool meets the bow wave 
pushed out ahead of the wheel. Instead, one might impose
conditions based on conservation of mass and force balance,
assuming that the bow wave acts like a jump discontinuity.
If $x=x_L$ denotes the position of the jump,
force balance again demands that the pressure
returns to atmospheric values:
\be
p(x_L,z,t)=0.
\label{bcforcL0}
\ee
Conservation of mass, on the other hand, indicates that
the normal velocity $V_L$  to the bow wave must satisfy
\be
(h_L - h_{in}) V_L % x_{Lt}
=
\left[{\bf Q}_L - \left(\begin{array}{c} U h_{in} \\ 0 \end{array}\right)
  \right] {\bf \cdot \; \hat{n}}
\label{1.10}
\ee
where ${\bf \hat{n}}$ denotes the normal to the bow wave in
the $(x,z)-$plane and ${\bf Q}_L={\bf Q}(x_L,z,t)$ is the local gap-average
flux, with
\be
   {\bf Q} =
   \left(\begin{array}{c} U h \\ 0 \end{array}\right)
   - \frac{h^3}{12\mu}
   \left(\begin{array}{c} p_x \\ p_z \end{array}\right)
   .
   \label{bcflux0}
\ee
This condition, however, implies that the bow-wave position $x_L$
is a function of both time $t$ and position $z$.
There is also no guarantee that some important physical effects
are thereby missed at the bow wave. 

Similarly, we must impose jump conditions at the rear of the gap
where lubrication theory fails to account for fluid mechanics
arising where the meniscus splits.
Indeed, experiments reveal complicated small-scale filamentation
due to either pressures becoming very low, or the printer's
instability, or both. Following common practice in
the bearing literature
\cite{pearson60,taylor63,dowson,coyle,coyle97},
we assume that pressures
return to atmospheric values with zero normal gradient:
\be
p = {\bf \hat{n} \; \cdot} \nabla p = 0
\quad {\rm at} \quad x=x_R(z,t),
\ee
or (equivalently)
\be
p(x_R,z,t)=p_x(x_R,z,t)=0.
\label{1.14}
\ee
Once more, these conditions render the film-splitting location
dependent on $z$, and do not obviously capture all the
dominant physical effects.

Note that the conditions \eqref{1.14} do not strictly imply
that film splitting arises at $x=x_R$, but that the gap for
$x>x_R$ is filled by a mixture of phases. That is, in front
of the lubrication zone, there is a potentially complicated
mix of the viscous liquid with either ambient air or vapour
(if the pressure becomes sufficiently low that cavitation arises).
Somewhat further to the right, when any filaments have broken, the
implication is that the liquid layer has
effectively split to form two roughly equal films on the wheel and track.

To side step the awkward issues at the front and back,
we first note that there are
two important limits of the problem in which difficulties are
suppressed: first,
for an infinitely wide wheel, as considered in \S \ref{wide},
the fluid flow becomes two-dimensional and independent of $z$,
with the conditions at the sides in \eqref{bcsides0} becoming
irrevelant. Even if the wheel is finite, but relatively wide,
one expects flow to remain largely two dimensional
and \eqref{bcsides0} to introduce corrections to a largely
$z-$independent pressure distribution only over boundary layers
near the wheel edges. Those boundary-layer corrections are not likely
to affect global mass or force balances. Thus, $x_L$ and $x_R$
become only functions of time.

Second, when the wheel is relatively narrow, variations across the
wheel are stronger than those in the direction of rolling
throughout most of the lubrication zone. We exploit this feature
in \S \ref{narrow} to derive solutions for narrow wheels.
One feature of those solutions is that, because some of the
derivatives in the rolling direction are neglected in comparison
to $z-$derivatives, it is no longer possible to impose the
force balance condition \eqref{bcforcL0} (the meniscus conditions
in \eqref{1.14} turn out to be satisfied more straightforwardly).
This failure
reflects the presence of a boundary layer at the bow wave with a
thickness of order the wheel width. This implies that the bow wave itself
has a shape over a similar scale in $x$. Nevertheless,
that boundary
layer again plays a minor role in the global mass and force balances,
again indicating that $x_L$ and $x_R$ are mostly
functions of time over characteristic length scales in the rolling direction.

To bridge the gap between these two limits (as in \S \ref{inbetween}),
we adopt a convenient approximation in which we simply assume that both
$x_L$ and $x_R$ are independent of $z$. This demands that we cannot
impose the full details of the boundary conditions in (\ref{1.10})
and (\ref{1.14}). Instead,
we impose only the pressure conditions $p(x_L,z,t)=p(x_R,z,t)=0$.
To ensure a consistent solution in which $x_L$ and $x_R$ can be taken
to be independent of $z$, we then further demand that the flux condition
in \eqref{bcflux0} and $p_x(x_R,z,t)=0$ are satisfied only in a
wheel-averaged sense. More details of this procedure are given in
\S \ref{inbetween} and the Appendix, where we also examine the fidelity of
the approximation.

\subsection{Dimensionless model}

To place the model into a more convenient dimensionless form,
we now introduce the scalings,
\be
(\xi,\zeta,\xi_L,\xi_R,W) = \frac{(x,z,x_L,x_R,\WW)}{\sqrt{Rh_*}},
\qquad
(\eta,\eta_0) = \frac{(h,h_0)}{h_*}
,
\ee
\be
\cP = \frac{h_* p \sqrt{Rh_*}}{12\mu U R}
, \qquad
\hat{t} = \frac{Ut}{\sqrt{Rh_*}}
, 
\ee
where $h_*$ is a characteristic measure of the minimum gap.
After dropping the hat decoration on $t$,
the Reynolds equation becomes
\be
\left( \xi \eta_{0t} + \eta - \eta^3 \cP_\xi \right)_\xi
-
\left(\eta^3\cP_\zeta \right)_\zeta
= 0,
\qquad
\eta = \eta_0(t) + \half \xi^2
.
\label{2.3}
\ee
The boundary conditions become
\begin{align}
\cP &= 0, \quad {\rm at} \ \zeta=\pm\half W, \\
\cP &= 0, \quad
(\eta-\eta_{in}) \xi_{Lt} = \eta-\eta_{in}
- \eta^3\left(\cP_\xi - \xi_{L\zeta} \cP_\zeta\right)
\quad {\rm at} \ \xi=\xi_L, \label{2.5} \\
\cP &= 0, \quad \cP_\xi=0,
\quad {\rm at} \ \xi=\xi_R.\label{2.6}
\end{align}

The dimensionless equation of motion of the wheel is
\be
M \ddot\eta_0 = \cL(\xi_L,\xi_R,\eta_0) - \cL_0
\label{2.7}
\ee
where the dimensionless lubrication force per unit width is
\be
\cL(\xi_L,\xi_R,\eta_0) = \frac{1}{W}
\int_{-W/2}^{W/2} \int_{\xi_L}^{\xi_R}  \cP(\xi,\zeta,t)\; \dd \xi \dd \zeta
\ee
and
\be
M = \frac{h_* \MM U}{12 \mu R^2 \WW}
\qquad {\rm and} \qquad
\cL_0 = \frac{h_* L}{12\mu U R}
\label{ML0}
\ee
are the dimensionless wheel mass and load. 
Assuming that both dimensionless groups are order one, we may
rewrite the mass parameter in \eqref{ML0} as
\be
M = \frac{\pi \rho_w \cL_0 R U^2 }{L}
.
\ee
Assuming that $\rho_w=10^4$ kg/m, $R=0.1$m and $U=1$m/s,
the loads used in the experiments ($L=10^4 $N/m)
translate to a mass parameter $M$ of order a tenth.

In practice, it is the balance between the load and
and the lubrication lift force that
dictates the minimim gap, at least in steady planing.
Therefore, one option for
fixing the scale $h_*$ is to take $\cL_0=1$, or
$h_0=12\mu U R / L$. This, however, leaves $\eta_0$ as an unknown
parameter in the problem that must be found as part of the solution.
A more convenient choice in that setting
is to demand that $\eta_0=1$ in the steady state;
{\it i.e.} $h_*$ is the minimum gap for steady planing.
This implies the presence of an order-one load parameter
\be
\cL_0 \equiv \cL(\Xi_L,\Xi_R,1),
\label{L0}
\ee
where $\xi_{L,R} = \Xi_{L,R}$ denote the right and left edges
of the lubrication zone in steady planing.

However, an awkward flaw in this choice for $h_*$
is that it becomes difficult to compare solutions
for the same load ({\it i.e.} wheel and axis system)
but different incoming fluxes $\eta_{in}$, or gauge
the effect of the wheel's width $W$ at fixed
load and flux. In view of this issue, we adopt the following
practice for setting $h_*$: we first pick a reference
state corresponding to a given load and flux. We then use
the minimum gap of the planing solution for this state
to fix $h_*$. This choice prescribes $\cL_0$. It also
proves convenient to fix $\Xi_L$ for the reference solution,
rather than the flux, and then compute the corresponding
$\eta_{in}$ with that solution in hand. Once $\cL_0$ and
$\eta_{in}$ have been set in this manner, we then
vary $\eta_{in}$ or $W$ to explore the effect of varying the flux
or wheel width.

\section{Dynamics of an infinitely wide wheel}\label{wide}

Assuming that the wheel is infinitely wide, $W\gg1$, we
discard any dependence on $\zeta$ and arrive at the
much simpler problem for the pressure,
\be
\begin{aligned}
( \xi \dot\eta_{0} &+ \eta - \eta^3 \cP_\xi )_\xi = 0,\\
\cP &= 0, 
\quad {\rm at} \ \xi=\xi_L, \\
\cP &= 0, \quad \cP_\xi=0,
\quad {\rm at} \ \xi=\xi_R.
\end{aligned}
\ee
An integral of the first relation leads to
\be
\cP_\xi = 
\frac{1}{\eta^3}
[(\xi-\xi_R) \dot\eta_{0} + \eta - \eta_R] .
\ee
A second integral then implies that
\be
I_2 - (\eta_R + \xi_R \dot\eta_0) I_3 =
\half \dot\eta_0 (\eta_R^{-2} - \eta_L^{-2})
,\label{constraint}
\ee
where
\be
I_j = \int_{\xi_L}^{\xi_R} \frac{\dd\xi}{(\eta_0+\xi^2/2)^j}
\ee
(analytical expressions for $I_2$ and $I_3$ are easily
provided). We further have
\be
  \cL \equiv -\int_{\xi_L}^{\xi_R} \xi \cP_\xi \; \dd\xi 
  = \half \eta_R^{-1} \eta_L^{-2} ( \eta_R-\eta_L)^2
  - \dot\eta_0 \left[ \half \xi_R(\eta_R^{-2} - \eta_L^{-2})
    + 2(I_2-\eta_0 I_3)\right] .
  \label{forc}
\ee
The problem now boils down to solving
the coupled ordinary differential equations,
\be
\begin{aligned}
M \ddot\eta_0 & = \cL(\xi_L,\xi_R,\eta_0) - \cL_0
,\\
\dot\xi_{L} & =  \frac{\eta_R-\eta_{in}+(\xi_R-\xi_L)\dot\eta_0}{\eta_L-\eta_{in}}
,
\end{aligned}
%\qquad
%
\label{odes}
\ee
(following from \eqref{2.5} and \eqref{2.7}),
in conjunction with the constraint in \eqref{constraint}
and force law \eqref{forc}.

\begin{figure}
\begin{center}
\includegraphics[width=.8\linewidth]{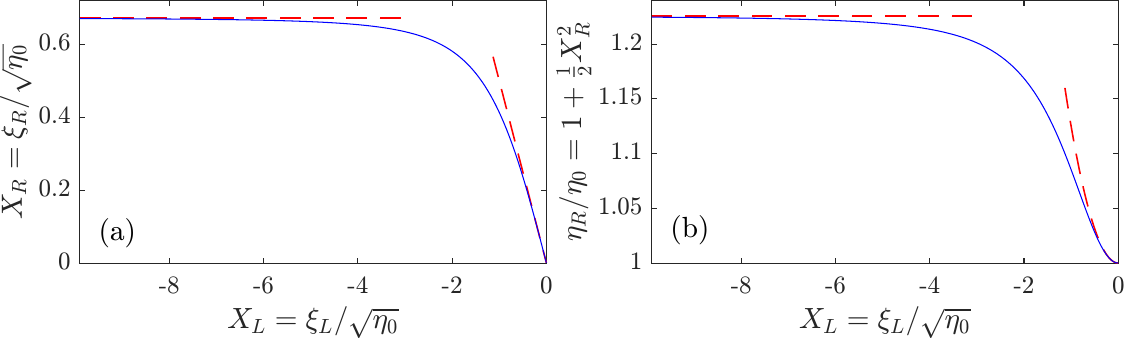}
\end{center}
\caption{\label{zRcomp}
  The implications of the constraint \eqref{constraint},
  plotting (a) $X_R$ and (b) $\eta_R/\eta_0=1+\frac12X_R^2$
  against $X_L$. The dashed lines show the limiting
  behaviours for $X_L\to-\infty$
  ($X_R\to 0.6719$ and $\eta_R\to1.2257\eta_0$)
  and $X_L\to0$
  ($X_R\to -\frac12X_L$ and $\eta_R\to1+\frac18X_L^2$).
}
\end{figure}

The constraint \eqref{constraint} can be rewritten
as the relation,
\be
\int_{X_L}^{X_R} \frac{(\chi^2-X_R^2)}{(2+\chi^2)^3}\dd\chi = 0,
\qquad
(X_{L},X_R)=\frac{(\xi_{L},\xi_R)}{\sqrt{\eta_0}}
.\label{conjalt}
\ee
Since the bow-wave position satisfies its own evolution equation, we interpret
this relation as prescribing
$X_R$ in terms of $X_L$. That prescription
is illustrated in figure \ref{zRcomp}.

\subsection{Steady planing versus flooding}\label{steadyflood}

Note that $X_R$ and $\eta_R/\eta_0=1+\frac12X_R^2$
have finite upper limits, reached for $X_L\to-\infty$.
This is significant because the steady planing state,
with $\eta_0$ and $(\xi_L,\xi_R)=(\Xi_L,\Xi_R)$ all constant, must satisfy
\be
\eta_R = \eta_{in}
\qquad {\rm and} \qquad
\cL_0=\cL(\Xi_L,\Xi_R,\eta_0) = \frac{(\eta_L^2-\eta_R)^2}{2\eta_R\eta_L^2}.
\label{stead}
\ee
These relations can be combined to find that
\be
\frac{\eta_R}{\eta_L} = \frac{2+X_R^2}{2+X_L^2} =
\sqrt{1 - \sqrt{2\eta_{in}\cL_0}},
\label{googoo}
\ee
which, in conjunction with \eqref{conjalt}, provides an
algebraic problem to solve for $(X_L,X_R)$. Returning to \eqref{stead},
one then arrives at the planing solution $(\eta_0,\Xi_L,\Xi_R)$.

However, \eqref{googoo} also highlights how
steady planing can only arise for $\eta_{in}< (2\cL_0)^{-1}$.
Larger fluxes necessarily lead to unsteady states, or flooding solutions,
in which the bow wave must continue to grow with time.
Physically, the flooding states arise because the pressure underneath
the narrowest parts of the gap becomes insensitive to the
bow-wave position when $|\Xi_L|\gg1$, and selects the minimum
gap according to the load. But this limiting gap can only
support a certain flux, so incoming pools with greater depth
choke and backflow.
The existence of both steady planing and flooding solutions
for an infinitely wide wheel was noted by \cite{rollpool}.

For the flooding state, the back flow for $|\xi_L|\gg1$
is characterized by
\be
\eta_R = (2\cL_0)^{-1} \approx 1.2257\eta_0
\label{floodit1}
\ee
(the limits of \eqref{conjalt} and \eqref{stead}). The bow-wave
evolution equation then demands that
\be
(1+\half\xi_L^2)\dot\xi_L \sim \eta_R-\eta_{in},
\qquad {\rm or} \qquad
\xi_L \sim -[6(\eta_{in}-\eta_R)t]^{1/3}.
\label{floodit2}
\ee

\subsection{Initial condition and early-time dynamics}

More generally,
initial conditions are required for $[\eta_0(0),\dot\eta_0(0),\xi_L(0)]$
in the system \eqref{odes} when solved as an initial-value problem.
One option is to consider the dynamics just prior to the lift-off
of the wheel: when the wheel makes contact with the initial
pool (and ignoring the shape), there is no film splitting or filamentation
at the right edge of the fluid region. Instead, fluid is pushed both
right and left behind steep waves at both the bow and rear.
As the wheel does not move vertically at this stage, the fluid flux
driven underneath the wetted section of the wheel is constant.
Consequently, beginning from the initial moment of contact
at which $\xi_L = \xi_r = -\sqrt{2\eta_{in}}$, the waves evolve
according to
\be
(\eta_R-\eta_{in})\dot\xi_R =
(\eta_L-\eta_{in})\dot\xi_L ,
\ee
and equal areas of fluid are pushed backward and forwards at
each moment of time. The expansion of the wetted region continues
until the lift force reaches the load, at which moment
\be
\cL_0 = \cL(\xi_L,\xi_R,0) = \half \eta_R^{-1} \eta_L^{-2} ( \eta_R-\eta_L)^2;
\ee
the wheel then takes off. The edges of the lubrication zone can be
calculated accordingly; we denote these positions by
\be
   [\xi_R(0),\xi_L(0)] = [\xi_{R*},\xi_{L*}]
   .\label{xiics}
\ee

Unfortunately, as soon at the wheel takes off, the model in 
\eqref{constraint}-\eqref{odes} demands that the wave at the right edge
of the fluid is instantaneously transformed into a film splitting or
filamentation point with $\cP(\xi_R,t)=0$, with a mixed-phase
region to its right.
But if we continue to assume that $\dot\eta=0$,
then the position of the right-hand wave, $\xi_R(0)$, is not
consistent with the constraint. Conversely, if we instead assume
that the wheel suddenly takes off with a speed $\dot\eta_0(0)=\dot\eta_{0*}$
chosen to satisfy the constraint with those values of $\xi_R(0)=\xi_{R*}$
and $\xi_L(0)=\xi_{L*}$, there is a
sudden change in lift force that typically forces an initial
inertial adjustment. Nevertheless, one possible option for the
initial condition is the selection,
\be
   [\eta_0(0),\dot\eta_0(0),\xi_L(0)]=[\varepsilon,\dot\eta_{0*},\xi_{L*}]
   . \label{ics1}
\ee
Note that here we take $\eta_0(0)$ to be a small
positive constant $\varepsilon>0$,
to avoid any issues with taking a zero minimum gap in the model equations.

\begin{figure}%[htb!]
\begin{center}
\includegraphics[width=.85\linewidth]{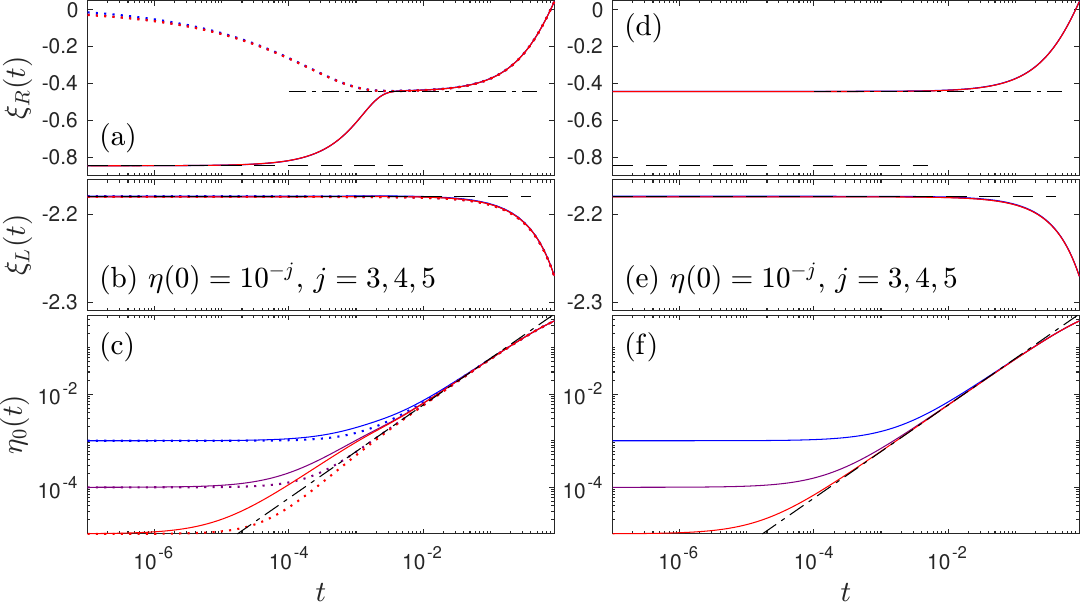}
\end{center}
\caption{\label{solini}
  Model solutions showing lift off, starting with
  $\eta_0(0)=\varepsilon=10^{-j}$, $j=3,4,5$ (colour coded, from blue to red),
  with $M=10^{-3}$ and
  adopting a flux $\eta_{in}$ and load $\cL_0$
  such that the steady minimim gap and bow-wave position are $\eta_0=1$ and
  $\xi_L=\Xi_L=-5$. In (a,b,c), the initial condition (indicated by dashed
  lines) are chosen so that the lift force
  due to the displaced pool to the left of the
  dry contact position reaches the load (equation (\ref{ics1})). In
  (d,e,f), the initial conditions
  are given by the quasi-steady solution as in \eqref{ics2}
  ($\xi_{Rs}$ and $\eta_0\sim\dot\eta_{0s}$ are shown by dot-dashed lines).
  The dotted lines in (a,b,c) show a further solution in which
  $\xi_R(0)$ is arbitrarily reset to the centre of the wheel
  and $\dot\eta_0(0)$ is adjusted to satisfy the constraint accordingly.
}
\end{figure}

Figure \ref{solini}(a,b,c) shows some solutions to this initial-value
problem, adopting three values for $\varepsilon$. Parameters are chosen
so that one anticipates a steady planing state with $(\eta_0,\Xi_L)=(1,-5)$,
and $M=10^{-3}$. Also shown
is another set of solutions (with the same choices for the parameters)
in which $\xi_{R}(0)$ is arbitrarily reset to zero, retaining
$\eta_{L}(0)=\eta_{L*}$ and
recomputing $\dot\eta_0(0)$ to satisfy the constraint.
The pairs of solutions highlight how the value of
$\varepsilon$ is not significant: the wheel lifts
up from the initially different positions, but then
all the pairs converge to a common solution after a short transient
involving an adjustment in $\eta_R(t)$.
The common solution is quasi-steady, in the sense that the
edges of the lubrication zone remain roughly in place with
the bow wave at its initial position; all the while,
the minimum gap grows steadily from its initially small value.

Because the quasi-steady state with $\xi_R=\xi_{Rs}$ and
  $\dot\eta_0  = \dot\eta_{0s} $,
the main balance in (\ref{odes}a) is
  \be
  \cL_0  \sim \cL(\xi_L,\xi_R,0)
  \qquad {\rm and} \qquad
  I_2 - (\eta_{Rs} + \xi_{Rs} \dot\eta_{0s}) I_3 =
  \half \dot\eta_{0s} [\eta_{Rs}^{-2} - \eta_L(0)^{-2}]
  ,
  \ee
  with
  \be
    \eta \sim \half \xi^2
  ,
  \qquad
  I_j \sim 2^j \int_{\xi_L}^{\xi_R} \frac{\dd\xi}{\xi^{2j}}
  \sim -\frac{2^j [\xi_{Rs}^{-2j+1}-\xi_L(0)^{-2j+1}]}{2j-1}
  .
  \ee
  These relations imply
  that $\xi_{Rs}$ and $\dot\eta_{0s}$ are approximately constant
  and dictated by the initial bow wave position $\xi_L(0)$.
  The resulting  predictions for $\xi_{Rs}$ and
  $\eta_0\sim\dot\eta_0 t = \dot\eta_{0s} t$
  are indicated by the dot-dashed lines in figure \ref{solini}.
  If $|\xi_L|\gg1$, some further reductions establish that
  \be
  \cL_0 \sim
  \xi_{Rs}^{-2} + \frac23 \dot\eta_{0s} \xi_{Rs}^{-3}
  \qquad {\rm and} \qquad
  0   \sim -\frac{8}{15}\xi_{Rs}^{-3}   - \frac25 \dot\eta_{0s} \xi_{Rs}^{-4}
  .
   \ee
   Hence
   \be
   \dot\eta_{0s} \sim - 4\cL_0^{-1/2}
   \qquad {\rm and} \qquad
   \xi_{Rs} \sim - \thir \cL_0^{-1/2}
   .
   \label{e38}
   \ee
   
   Figure \ref{solini}(d,e,f) show solutions starting from the
   alternative initial conditions implied by the quasti-steady
   solution:
   \be
      [\eta_0(0),\dot\eta_0(0),\xi_L(0)]=[\varepsilon,\dot\eta_{0s},\eta_{L*}]
.
   \label{ics2}
   \ee
   Because the initial transient is eliminated for these cases,
   \eqref{ics2} arguably provides the most natural initial
   conditions for the model. 
   Henceforth, we adopt
   \eqref{ics2}. Regardless of the choice of initial condition,
   however, the minimum gap
   $\eta_0(t)$ is expected to reach order unity values over times of
   order one.

   Note that the right edge of the
   lubrication zone always lies to the left of the minimum gap
   at the beginning of the computations. This position cannot
   therefore be interpreted as a position of film splitting
   until it migrates to the right of the minimum gap. Instead, we must
   assume that there is a mixed-phase region underneath the wheel.
   A better resolution of this awkward point demands a more detailed
   model for the right-hand edge of the lubrication zone.

\subsection{Later-time dynamics}

\begin{figure}[htb!]
\begin{center}
\includegraphics[height=.51\linewidth]{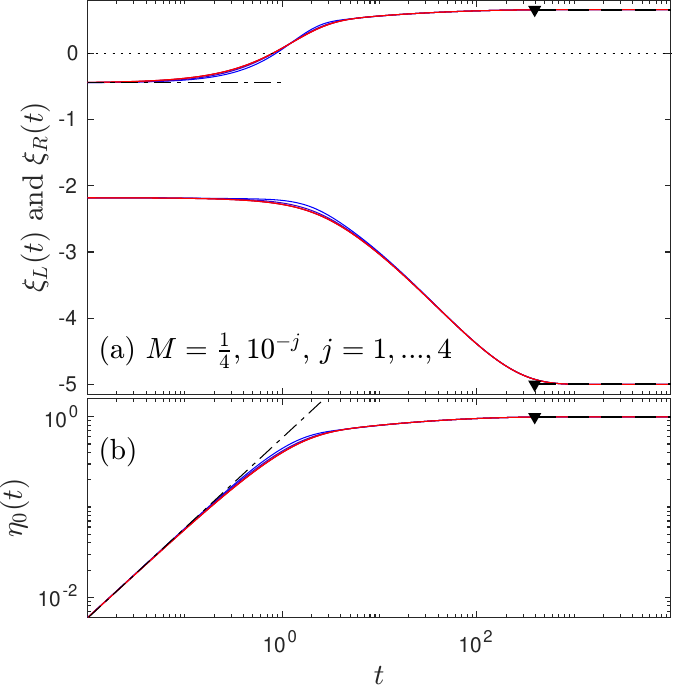}
\includegraphics[height=.51\linewidth]{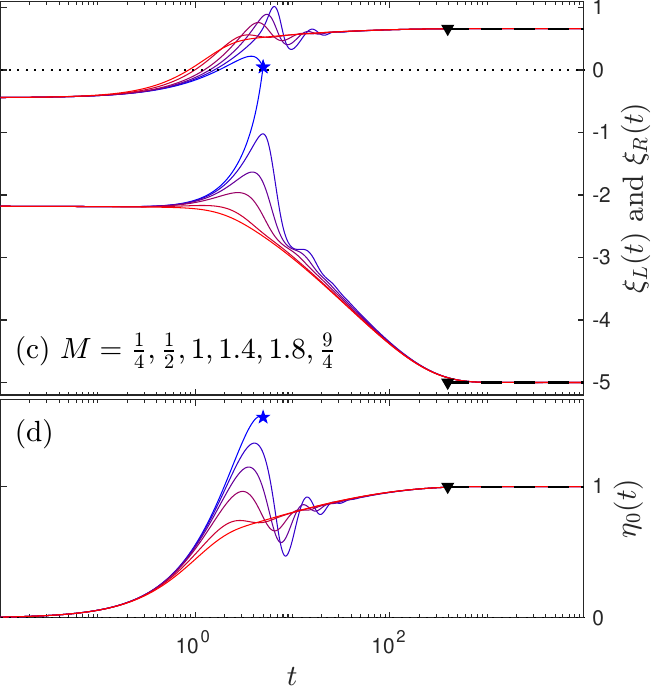}
\end{center}
\caption{\label{sold}
  Model solutions showing lift off, starting with \eqref{ics2} and
  $\eta_0(0)=\varepsilon=10^{-4}$, for
  a flux $\eta_{in}$ such that the steady bow-wave position is
  $\xi_L=\Xi_L=-5$. (a,c) Time series of the positions of the bow
  wave and lubrication front,
  $\xi_R(t)$ and $\xi_L(t)$, and (b,d) minimum gap $\eta_0(t)$.
  Solutions for different mass parameters are presented:
  (a,b) $M=10^{-j},\frac14$, $j=\{4,3,2,1\}$, and
  (c,d) $M=\frac14$, $\frac12$, 1, 2, $3.5$ and 5
  (in both cases, colour-coded, from red to blue).
  The dashed lines indicate the steady final planing state,
  with the triangle marking $t=3t_\infty = 3|\xi_{L\infty}|^5/(16\eta_{in}^2)$.
  The dot-dashed lines in (a,b) show $\xi_{Rs}$ and $\dot\eta_{0s}t$.
  The star indicates the time at which the lubrication zone
  for the solution with largest
  $M$ shrinks to a point.
}
\end{figure}

Figure \ref{sold} displays the later-time dynamics of solutions
starting with \eqref{ics2}, adopting different values
for the mass parameter $M$, but again taking $(\eta_0,\Xi_L)=(1,-5)$
for the steady planing state. In all cases,
the initial lift-up of the wheel takes place relatively quickly,
with the bow wave holding its initial position.
Wheel inertia plays little role for the lower values of $M$ (panels (a,b)),
with all the solutions collapsing close to one another.
Both the ascent of the wheel and the rightward extension
of the lubrication zone slow
once the gap opens towards the steady planing value.
The bow wave continues to build at this
stage, however, and only reaches its steady state over rather
longer times.
With large values of $M$, although the initial ascent is similar,
the inertia of the wheel first delays the ascent at later times,
and then causes the opening of the gap to overshoot and proceed into
a series of decaying oscillations. In fact, for the highest value of $M$,
the inertial overshoot is sufficiently extreme to prompt the
lubrication zone to shrink to a point, implying that the wheel
leaves contact with the viscous pool.

  The continued expansion of the bow wave at later times arises
  because the lubrication force is relatively insensitive
  to the bow-wave position. Consequently, once the minimum
  gap opens to near its steady planing value, there is
  only a minor further adjustment to the vertical
  position of the wheel. However, the
  residual mismatch between the incoming flux
  and that leaking underneath the wheel leads to the migration
  of the bow wave. In the solution pictured, this final expansion
  looks to take well over a hundred time units.
  Because this period is so long, the evolution takes place quasi-statically,
  with the $\dot\eta_0$ terms in \eqref{constraint}-\eqref{forc}
  playing little role and the wheel roughly in force balance.
  The bow wave then expands according to
  \be
  \dot\xi_L \sim \frac{\eta_R-\eta_{in}}{\eta_L-\eta_{in}}
  \label{e39}
  \ee
  with
  \be
%  \int_{z_L}^{z_R} \frac{z^2 - z_R^2}{(1+\half z^2)^3} \dd z\sim 0
%  \qquad {\rm and} \qquad
  \dot\eta_0 \sim \frac{(X_L^2-X_R^2)^2}{(2+X_L^2)^2(2+X_R^2)\cL_0}
  .
  \label{e40}
  \ee
  The constraint \eqref{conjalt} and \eqref{e40} now dictate $X_R$
  and $\eta_0$ in terms of $X_L$, 
  or, equivalently, $\xi_R=X_R\sqrt{\eta_0}$ and $\eta_0$ in terms of
  $\xi_L=X_L\sqrt{\eta_0}$. The remaining ODE \eqref{e39}, can then be solved
  for $\xi_L(t)$. However, a closer examination of the integrals in
  \eqref{e40} indicates that when $|\xi_L|$ is relatively large,
  the corrections to the constraint are smaller than those
  to the force balance by one order in $|\xi_L|^{-1}$. Consequently,
  we may use the final, steady value of $X_R$ as a convenient approximation
  of the constraint;
  {\it i.e.} $X_R\sim \sqrt{2(\eta_{in}-1)}$.
  Moreover, $\eta_L-\eta_{in}\sim\frac12 \xi_L^2$.
  Hence,
  \be
  \dot\xi_L \sim \frac{2\eta_{in}}{\xi_L^2}(\eta_0-1)
  .
  \ee
  In the approach to the steady state,
  $\eta_0\sim1+8\eta_{in}(\xi_L-\xi_{L\infty})/\xi_{L\infty}^3$
  if $(\xi_L,X_L)\to\xi_{L\infty}$,
  and we may deduce that
  \be
  \dot{\eta_0} \sim \frac{16\eta_{in}^2}{\xi_{L\infty}^5} (\eta_0-1).
  \ee
  The decay towards the final state therefore takes place over
  a relatively long timescale
  \be
  t_\infty = \frac{|\xi_{L\infty}|^5}{16\eta_{in}^2}
  ,
  \label{tinf}
  \ee
  when $|\xi_{L\infty}|\gg1$. For the
  examples in figure \ref{sold}, the timescale
  $t_\infty\approx131$.
  As also indicated in that figure, a rough indicator for when
  the bow wave moves close to its steady state is provided by the
  estimate $t\approx3t_\infty$.

\begin{figure}
\begin{center}
\includegraphics[width=.9\linewidth]{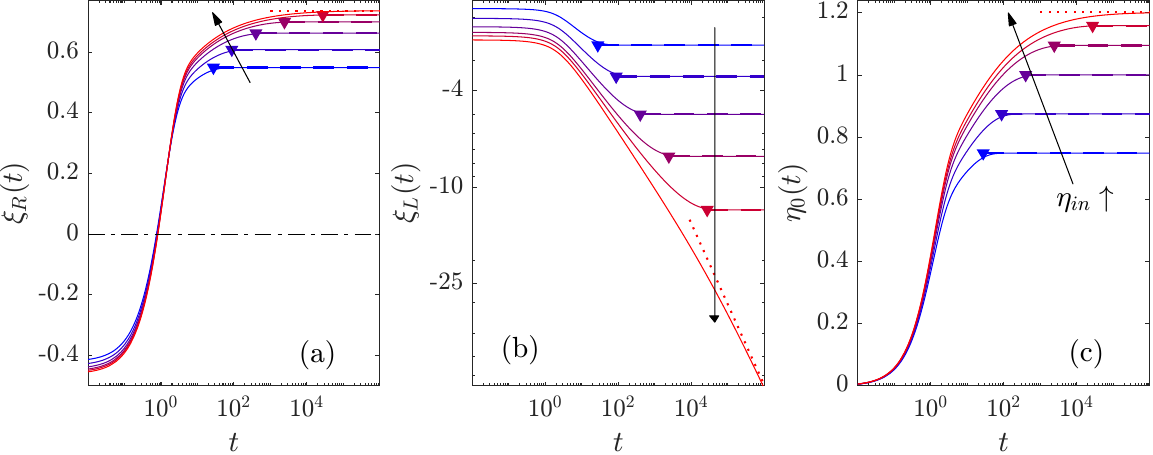}
\end{center}
\caption{\label{soldxx}
  Solutions to \eqref{odes} for $M=10^{-3}$, the initial
  conditions in \eqref{ics2}, and the same load parameter
  $\cL_0$ as in figure \ref{sold}. Six solutions are
  shown, corresponding to fluxes of
  $\eta_{in}=\{0.9,1.06,1.22,1.34,1.42,1.52\}$
  (colour-coded from blue to red).
  The dashed lines show the expected steady planing state
  (\S\ref{steadyflood}) with the triangle marking $t=3t_\infty$.
  The dotted lines show the predictions \eqref{floodit1} and
  \eqref{floodit2}.
}
\end{figure}

The choices for the parameters for the
solutions in figure \ref{sold} ensure that there are steady planing states.
To examine the transition to planing in more detail, figure \ref{soldxx}
shows further solutions in which we select
the initial conditions in \eqref{ics2} and the same values for
the load $\cL_0$ and $M$,  but then vary the flux $\eta_{in}$.
Six examples are shown. All but the case with the highest flux
lead to steady planing. Because the flux is now changing but the
load is fixed, the steady planing states have different minimum gaps
and bow-wave positions that match up with predictions from
\S \ref{steadyflood}. The time taken to reach these states
is again estimated by $3t_\infty$ (see the triangles in figure \ref{soldxx}).
The last example has no steady state because
$2\eta_{in}\cL_0 \approx 1.016>1$ for this case, implying
that a flooding solution is expected instead. Indeed, the bow
wave continues to move to the left for this example,
with the solution converging to
the predictions in \eqref{floodit1} and \eqref{floodit2}.

  \subsection{Touch down}

\begin{figure}
\begin{center}
\includegraphics[width=.65\linewidth]{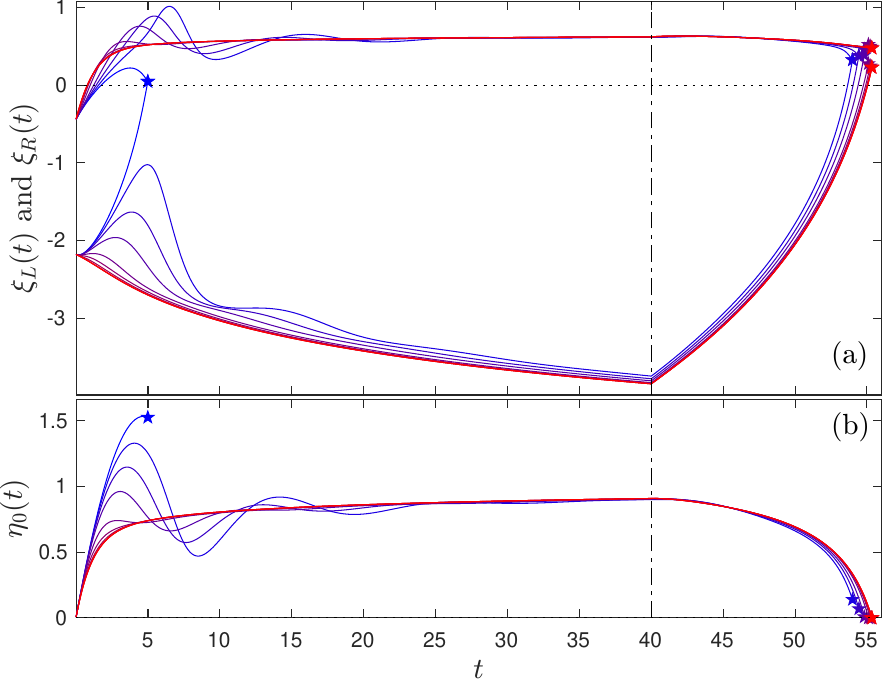}
\end{center}
\caption{\label{sold2}
  The model solutions of figure \ref{sold},
  but showing touch down when the incoming flux $\eta_{in}$ is
  turned off at $t=t_e=40$.
  The stars indicate the final times, at which
  the wheel either again makes contact with the track
  or fluid becomes flushed out of the gap with the wheel at finite
  height. (Again, the different solutions are colour-coded by
  $M$, with $M$ increasing from red to blue).
}
\end{figure}

  In the experiments, the pool has finite length. Consequently,
  in the model, the incoming flux $\eta_{in}$ must be turned off
  after a time,
  \be
  t_e = \frac{L_P}{\sqrt{Rh_*}}
  ,
  \ee
  where $L_P$ is the dimensional pool length.
  Figure \ref{sold2} reports this version of the initial-value problem
  for the same solutions shown in figure \ref{sold}, but now with
  $t_e=40$.
  The switch-off in the flux leads to an abrupt change to the late-time
  dynamics, with the bow wave sharply changing direction and the gap
  beginning to close. That closure is more gradual to begin
  with, again because the lubrication lift force is relatively
  insensitive to the bow-wave position, at least when $|\xi_L|$ is
  sufficiently large. However, once $|\xi_L|$ decreases to
  smaller values, the lift force is more sigificantly reduced
  and the gap closes faster. For lower values of $M$ (redder curves), the
  gap closes before the lubrication region shrinks to a point;
  with higher inertia (bluer curves), that region collapses before
  the minimum gap can close.
  Note that in all the cases shown, including the one terminated at
  small times with the highest $M$, the lubrication zone
  has migrated to the right of the minimum gap. There is
  therefore no divergence of the lubrication force for
  $\eta_0\to0$, which would otherwise happen to prevent
  solid-solid contact in finite time, as in other
  sedimentation problems \cite{brenner,contactI}.
  To summarize: the wheel either touches down continuously
  or loses contact with the fluid at finite height because
  the fluid becomes flushed out of the gap (no contact-line
  pinning effects having been included in the model).
  
\section{Narrow wheels, $W\ll1$}\label{narrow}

\subsection{Simplified model}

For a narrow wheel, the natural lengthscale for variations in $\zeta$
is not $\sqrt{Rh_*}$, but the width $\WW\ll\sqrt{Rh_*}$.
Moreover, the effective mass and load on the wheel ($M$ and $\cL_0$)
should be correspondingly smaller, given the use of $\sqrt{Rh_*}$
in defining them.
We therefore rescale,
\be
\zeta = W \hze, \qquad
[\cP,M,\cL_0] = W^2 [\hcP(\xi,\hze,t),\hM,\hcL_0]
\label{4.1}
,
\ee
and write
\begin{align}
(\xi \eta_{0t} + \eta - W^2 \eta^3 \hcP_\xi)_\xi - \eta^3 \hcP_{\hze\hze} = 0
\label{4.2a}
  , \\
  \hM\ddot\eta_0 = 2W^2 \int_0^{\frac12} \int_{\xi_L}^{\xi_R} \hcP \; \dd\xi\dd\hze
  - \hcL_0
\label{4.2b}
\end{align}
Moreover, given the narrowness of the wheel, we expect that
the rounding of the front and back of the lubrication zone to
take place over lengths in $\xi$ of order $W$. Hence we also set
\be
\xi_{R,L} = \Xi_{R,L}(t) + W \Delta_{R,L}(\hze,t).
\label{4.3}
\ee
The boundary conditions at the front and back now become
\begin{align}
  \hcP &= \hcP_\xi =0 \quad {\rm at} \ \xi=\Xi_R+W\Delta_R
  , \\
\hcP &= 0, \quad 
(\eta-\eta_{in}) (\Xi_{Lt}+W\Delta_{Lt}) = \eta-\eta_{in}
- W\eta^3\left(W\hcP_\xi - \Delta_{L\zeta} \hcP_\zeta\right)
\quad {\rm at} \ \xi=\Xi_L+W\Delta_L,
\label{4.4}
  \end{align}

Over the bulk of the lubrication zone, we may omit the
$O(W^2)$ terms in \eqref{4.2a} to find that
\be
\hcP \sim \frac{\eta_{0t}+\xi}{2\eta^3} \left(\hze^2-\quar\right),
\label{4.5}
\ee
in which case
\be
\hM \ddot\eta_0 = \frac{1}{24}
\left( \eta_R^{-2}-\eta_L^{-2} - 2\dot\eta_{0} I_3\right) - \hcL_0
.
\label{4.6}
\ee
To avoid the unphysical conclusion that $\dot\Xi_L=1$, 
\eqref{4.4} also implies that $\eta_L-\eta_{in}=O(W)$. Thus,
\be
\Xi_L = - \sqrt{2(\eta_{in}-\eta_0)}
.
\ee
In other words, when the wheel is narrow, the side flux is too strong
to allow the build-up of any appreciable bow wave beyond the waterline
of the incoming pool.
We may further take care of the right-hand pressure condition,
$\hcP=0$ at $\xi=\xi_R\sim\Xi_R$,
at leading order by demanding $\Xi_R \sim -\dot\eta_{0}$.
Hence,
\be
\hM \ddot\eta_0 = \frac{1}{24}
\left[ (\eta_0+\half\dot\eta_0^2)^{-2}-\eta_{in}^{-2} - 2\dot\eta_{0}
  \int_{-\sqrt{2(\eta_{in}-\eta_0)}}^{-\dot\eta_0} \frac{\dd\xi}{\eta^3} \right]
- \hcL_0
.
\label{node}
\ee

Assuming that inertial effects remain small,
the initial condition for \eqref{node} can be taken to be
\be
   [\eta_0(0),\dot\eta_0(0)]=[\varepsilon,\dot\eta_{0s}],
   \label{nice}
\ee
where $\varepsilon\ll1$ (practically we use $\varepsilon=10^{-3}$)
and
\be
\frac15 \dot\eta_{0s}^{-4}-\eta_{in}^{-2} + \frac{16}{5}\dot\eta_{0s}
  (2\eta_{in})^{-5/2}
= 24 \hcL_0
\label{nicer}
\ee
dictates the take-off speed $\dot\eta_{0s}$
of the wheel (for which the lift force balances the load).
The corresponding initial edges for the lubrication zone are
$(\Xi_L,\Xi_R)=(-\sqrt{2\eta_{in}},-\dot\eta_{0s})$.

Note that the pressure distribution in \eqref{4.5} does not
satisfy the remaining boundary conditions, $\hcP=0$ at $\xi\sim\Xi_L$
and $\hcP_\xi=0$ at $\xi\sim\Xi_R$. This points to the presence
of additional boundary layers at the front and back of the lubrication
zone over which $\hcP$ becomes adjusted to eliminate these discrepancies.
The scale in $\xi$ of these boundary layers is $O(W)$, thereby
allowing the derivatives in the rolling direction to re-enter
the main balances.
%A summary of both boundary layer problems is given in Appendix...

\subsection{Sample solutions}

\begin{figure}
\begin{center}
\includegraphics[width=.9\linewidth]{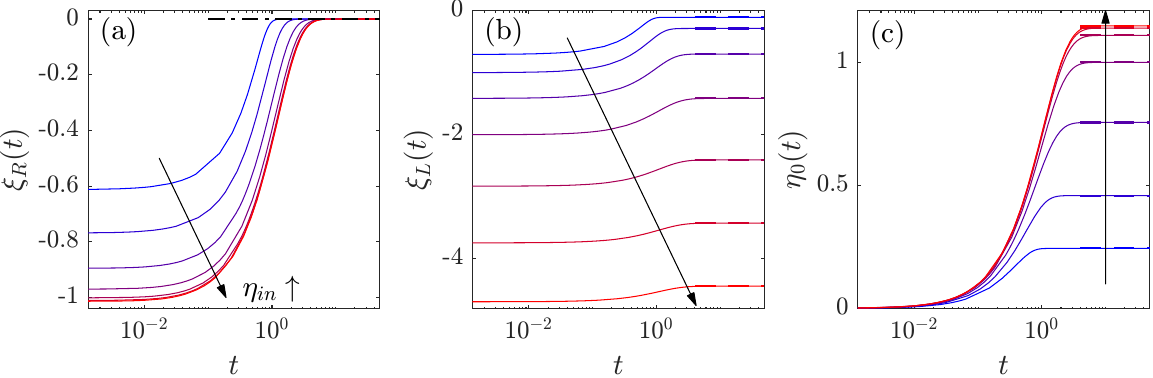}
\end{center}
\caption{\label{nl}
  Narrow-wheel solutions to \eqref{node} for varying flux
  ($\eta_{in}=\{\frac14,\frac12,1,2,4,7,11\}$,
  colour-coded from blue to red),
  with the load set such that steady planing with $\eta_{in}=2$
  has a minimum gap of unity. The dashed lines show the steady states
  from \eqref{nss}.
  ($\eta_0(0)\equiv\varepsilon=\hM=10^{-3}$).
}
\end{figure}

Equation \eqref{node} admits steady planing solutions with
\be
\eta_0 = \left[24\hcL_0 + \eta_{in}^{-2}\right]^{-\frac12},
\qquad \Xi_R=0, \qquad \Xi_L=-\sqrt{2(\eta_{in}-\eta_0)}.
\label{nss}
\ee
Figure \ref{nl} illustrates 
the progress of solutions to these steady states from the initial
condition \eqref{nice}, for several values of the incoming flux.
The inertia parameter is taken to be $\hM=10^{-3}$,
and the load is fixed by demanding that the case with flux $\eta_{in}=2$
has a steady planing state with a minimum gap of unity.

For all the cases displayed, the lubrication zone begins to the left
of the minimum gap. Both borders then move to the right
until $\Xi_R\to0$ ($\dot\eta_0\to0$) and the wheel reaches steady state.
The motion of the bow wave
is opposite to that for an infinitely wide wheel, in which
$\xi_L$ always expands to the left.
This feature arises precisely because the side flux
prevents the bow wave from building up above the waterline of the
incoming pool, and so the ascent of the wheel forces $\Xi_L$ to
move to the right. For large incoming flux, the solutions for the
minimum gap and $\Xi_R$ again converge to a bow-wave-independent
limit. This time, however, the bow-wave position $\Xi_L$ remains
close to its initial value, $\Xi_L\approx-\sqrt{2\eta_{in}}$,
and there are no flooding states, which
become eliminated by side flux.
The minimum gap in this limit is $\eta_0\approx (24\hcL_0)^{-1/2}$.

\begin{figure}
\begin{center}
\includegraphics[width=.9\linewidth]{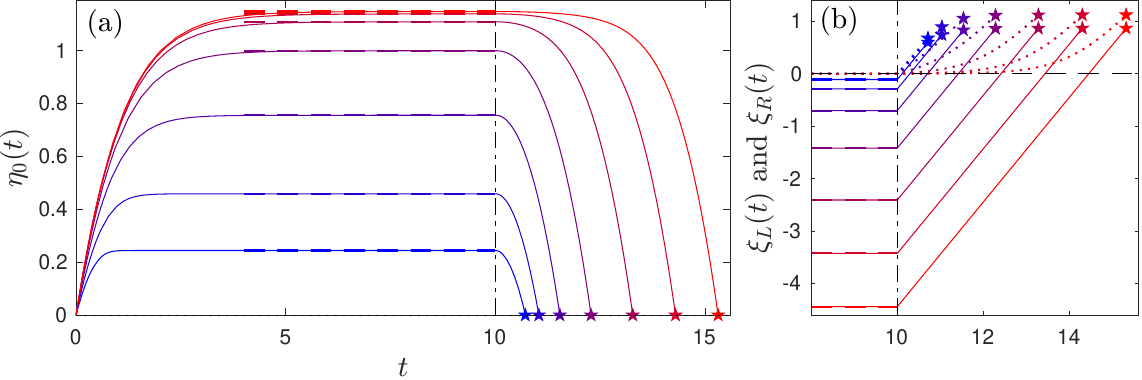}
\end{center}
\caption{\label{nl2}
  The model solutions of figure \ref{nl},
  but showing touch down when the incoming flux $\eta_{in}$ is
  turned off at $t=t_e=10$. The dashed lines show
  the steady planing solutions and
  the stars indicate the final times, at which
  the wheel makes contact with the track.
  In (b), the dotted lines show $\xi_R=\Xi_R$.
}
\end{figure}

When the pool ends and the incoming flux is turned off, it is no
longer possible to assume that $\eta_L\sim\eta_{in}$. Instead,
the bow wave must be carried back underneath the wheel, with
\be
\dot\Xi_L\sim1+O(W), \qquad {\rm or} \qquad \Xi_L = \Xi_L(t_e) + t - t_e.
\label{enar}
\ee
Similarly, we must then return to \eqref{4.6}, exploiting
this result for $\eta_L=\eta_0+\half\Xi_L^2$ and
taking $\eta_R\sim\eta_0 + \frac12\dot\eta_0^2$.
Figure \ref{nl2} shows
computations repeating those in figure \ref{nl},
but turning off the flux at $t=t_e=10$. In these examples,
with relatively low inertia, the wheel touches down
with a finite lubrication gap to the right of the
contact point.

Note that \eqref{enar} predicts that the time taken to touch down is 
nearly equal to that required for the bow wave to traverse the
original lubrication zone ({\it i.e.} $t-t_e\sim |\Xi_L(t_e)|$ in
our dimensionless notation). This is unlike the infinitely wide wheel,
for which the touch-down time is more prolonged (see figure \ref{sold2}),
owing to the pressure-driven back flow from the minimum gap to the
bow wave.

\section{Finite width}\label{inbetween}
 
%Our strategy for solving the two-dimensional Reynolds equation with unknown
%front and back positions relies on the assumption that both
%of these boundaries are nearly straight, an approximation that we interrogate
%{\it a posteriori}. 
%In doing so, the edge positions $\Xi_L$ and $\Xi_R$ are suitably chosen
%and 
%are set to one side.

To bridge between the limits of an infinitely wide or narrow
wheel, we adopt the convenient approximation that the
bow wave and front edge of the lubrication zone are straight.
This further demands that we set to one side
two of the boundary conditions in (\ref{2.5})-(\ref{2.6});
these conditions can only be satisfied in a wheel averaged
sense. More specifically, we fully retain the pressure conditions
$\cP(\xi_L,\zeta,t)=\cP(\xi_R,\zeta,t)$. Then, to fix
$\xi_R(t)$ and provide an evolution equation for the bow wave,
now at $\xi=\xi_L(t)$,
we demand
\begin{align}
\overline\cP_\xi &=0 \quad {\rm at} \ \xi=\xi_R, \label{2.6a} \\
(\eta-\eta_{in}) \xi_{Lt} &= \eta-\eta_{in}
- \eta^3 \overline\cP_\xi 
\quad {\rm at} \ \xi=\xi_L, \label{2.5a} 
\end{align}
where the overline denotes an average over the wheel:
\be
\overline{(...)} \equiv \frac{1}{W} \int_{-W/2}^{W/2} (...) \; \dd\zeta.
\ee

The Reynolds equation \eqref{2.3} must now be solved
on the rectangular domain,
$\xi_L(t) < \xi < \xi_R(t)$ and $-\half W < \zeta < \half W$.
At each moment of time, the minimum gap $\eta_0(t)$
and bow wave position $\xi_L(t)$ follow from having
time integrated the evolution equations in
\eqref{2.5a} and \eqref{2.7}. Evolving both forwards in time
then demands that we solve the Reynolds equation for the
pressure distribution, which dictates the lift force
$\cL(\xi_L,\xi_R,\eta)$ as well prescribing
the right-hand edge position $\xi_R(t)$ by enforcing
\eqref{2.6a}. We accomplish this task by adopting a trial
value for $\xi_R$, solving the Reynolds equation
subject to $\cP(\xi_R,\zeta,t)=0$, and then iteratively
updating $\xi_R(t)$ until \eqref{2.5a} is satisfied.

\subsection{Solution of the Reynolds equation}

The solution of the Reynolds equation 
over a known rectangular domain in $(\xi,\zeta)$ can be accomplished 
by separation of variables: we put
\be
\cP(\xi,\zeta,t) = \Pi(\xi,t) + \Phi(\xi,\zeta,t), 
\ee
where
\be
\left(\eta_{0t} \xi + \eta - \eta^3 \Pi_\xi\right)\xi = 0
\qquad {\rm and} \qquad
\left(\eta^3 \Phi_\xi\right)_\xi +
\left(\eta^3 \Phi_\zeta\right)_\zeta = 0
,
\label{3.2}
\ee
with
\be
\Pi(\xi_L,t) = \Pi(\xi_R,t) = \Phi(\xi_L,\zeta,t) =
\Phi(\xi_R,\zeta,t) = 0
\label{3.3}
\ee
and
\be
\Phi(\xi,\pm \half W,t) = -\Pi(\xi,t).
\ee
The function $\Pi(\xi,t)$ is given by
\be
\Pi =
\cI_2(\xi) - \frac{\cI_2(\xi_R)}{\cI_3(\xi_R)} \cI_3(\xi)
+ \frac{1}{2} \dot\eta_0 \left[ \eta_L^{-2}-\eta^{-2} +
  \frac{\cI_2(\xi_R)}{\cI_3(\xi_R)} (\eta_L^{-2}-\eta_R^{-2})
  \right]
,\qquad
\cI_j(\xi) = \int_{\xi_L}^\xi \frac{\dd\xi}{\eta^j}
\ee
(suppressing the dependence on $t$ for the integral functions
${\cal I}_j$).

The series solution of the partial differential equation in (\ref{3.2}) 
is then given by
\be
\Phi(\xi,\zeta,t) = - \sum_{j=1}^\infty 
\frac{c_j \phi_j \cosh \lambda_j \zeta}{\cosh \half\lambda_jW},
\label{3.6}
\ee
where
\be
c_j(t) = \int_{\Xi_L}^{\Xi_R} \Pi(\xi,t) \phi_j(\xi,t) \eta^3 \dd\xi
\label{3.7}
\ee
and the eigenfunctions $phi_j(\xi)$ solve the Sturm-Liouville problem,
\be
(\eta^3 \phi_j')' + \lambda_j^2 \eta^3 \phi_j = 0,
\quad \phi_j(\xi_L,t)=\phi_j(\xi_R,t)=0,
\quad \int_{\xi_L}^{\Xi_R} [\phi_j(\xi,t)]^2 \eta^3 \dd\xi = 1
\label{3.8}
\ee
(in which the time dependence enters entirely parametrically
through $\eta_0(t)$).

\begin{figure}
\begin{center}
\includegraphics[width=\linewidth]{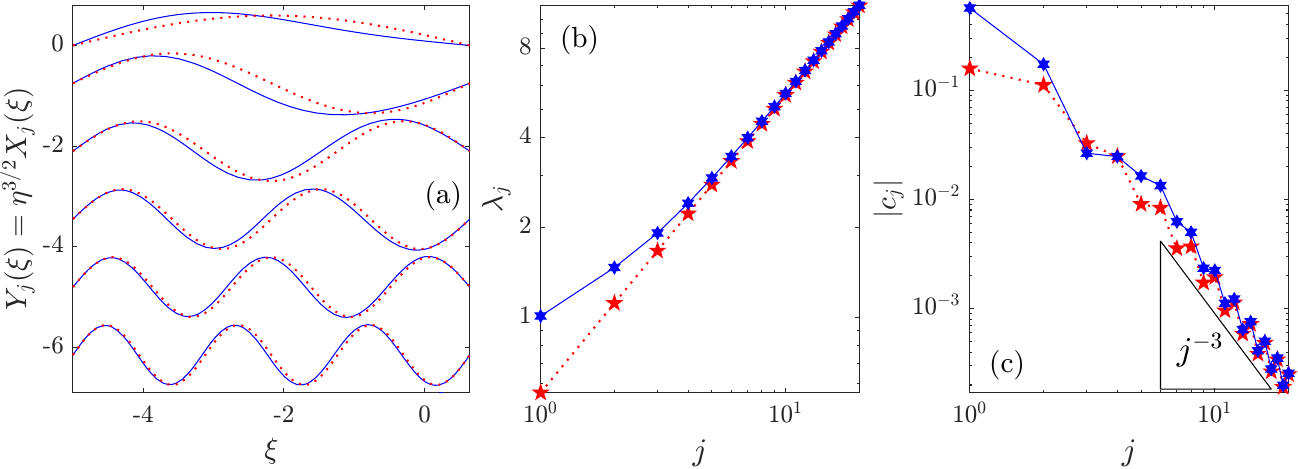}
\end{center}
\caption{\label{WKBplot}
  (a) The first six eigenfunctions of the
  Sturm-Liouville problem for $(\xi_L,\xi_R,\eta_0)=(-5,0.642,1)$.
  To remove the main variation of the amplitude of the modes,
  we plot $\eta^{-3/2}\phi_j(\xi)$ against $\xi$.
  The (corresponding) first twenty (b) eigenvalues $\lambda_j$ and (c)
  expansion coefficients $c_j$, plotted against $j$. The
  (red) dotted lines show the WKB approximation in \eqref{WKB}.
}
\end{figure}

The Sturm-Liouville eigensolutions $\{\lambda_j,\phi_j\}$
can be found by numerically solving (\ref{3.8}).
Sample low-order eigenfunctions are displayed in figure
\ref{WKBplot}, adopting $(\xi_L,\xi_R,\eta_0)=(-5,0.642,1)$.
Also plotted are the lowest eigenvalues
$\lambda_j$ and expansion coefficients $c_j$, along with the
useful WKB approximation, 
\be
\lambda_j \sim \frac{\pi j}{\xi_R-\xi_L},
\quad
\phi_j \sim \eta^{-3/2} \sqrt{\frac{2}{\xi_R-\xi_L}} \sin \lambda_j (\xi-\xi_L),
\quad
c_j \sim \frac{\sqrt2 (\xi_R-\xi_L)^{5/2}}{\pi^3 j^3}\left[
  (-1)^j\frac{\xi_R}{\eta_R^{3/2}} - \frac{\xi_L}{\eta_L^{3/2}} \right]
.
\label{WKB}
\ee
This WKB solution highlights how the coefficients $c_j$ decay like the
power law $j^{-3}$ for $j\gg1$. 

\begin{figure}%[htb!]
  \begin{center}
\includegraphics[width=\linewidth]{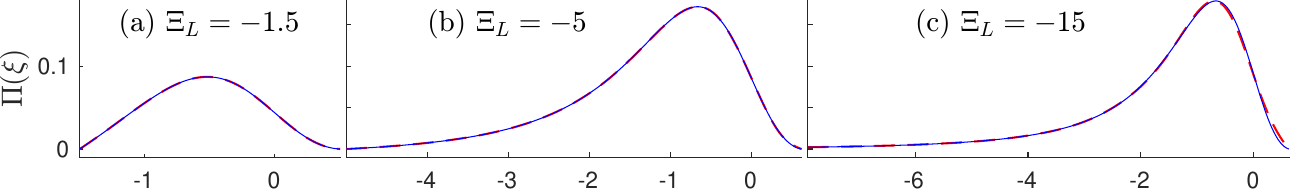}
\end{center}
\caption{\label{Qplot0}
  The representation of $\Pi(\xi)$ (solid blue) by the
  series of Sturm-Liouville eigenfunctions
  truncated to 20 terms (dashed red).
  Only part of the domain is shown in (c), the solution
  becoming small further to the left.
}
\end{figure}

Practically, we find that
truncating the Sturm-Liouville series  in (\ref{3.6}) at $j=J=32$ suffices
for accurate numerical results, although we often take higher values
to ensure this is the case. These conclusions can be justified by noting
that truncations of the series are least accurate
along the edges of the wheel
(the factor $\cosh\lambda_j\zeta/\cosh\frac12\lambda_jW$
decays least quickly with $j$ when $\zeta=\pm\frac12 W$).
Consequently, the accuracy of a particular truncation can be gauged
by examining the degree to which the series
$\sum_j c_j \phi_j(\xi)$ approximates $\Pi(\xi)$.
This is illustrated in figure \ref{Qplot0}
for three values of $\xi_L$, and taking $\eta_0=1$.
For the cases shown, twenty terms of the
series are sufficient to adequately reproduce the shape of
$\Pi(\xi)$. For larger values of $|\xi_L|$ that those shown,
the increasing localization of $\Pi(\xi)$ to the narrowest section
demands the inclusion of more terms of the series.

\subsection{Steady planing solutions}

\begin{figure}%[htb!]
  \begin{center}
    (a) Pressure distributions
\includegraphics[width=\linewidth]{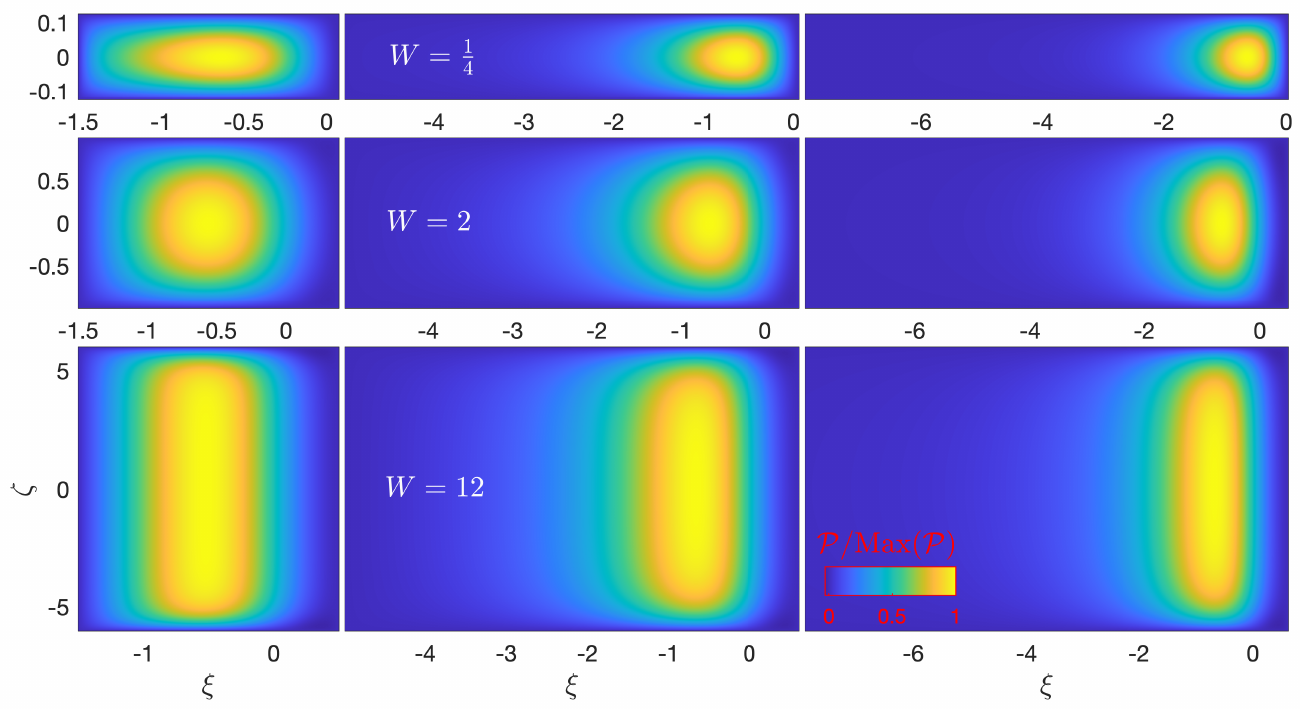}
    (b) Midline pressures \\
\includegraphics[width=.8\linewidth]{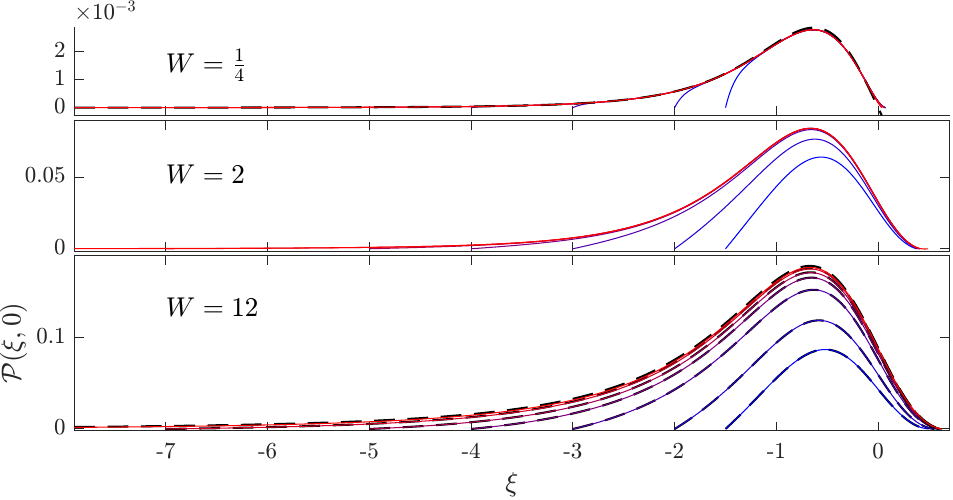}
\end{center}
\caption{\label{Qplot}
  (a) Numerical solutions of the Reynolds equation
  for $\Xi_L=-1.5$, $-5$ and $-15$ (left to right)
  and $W=\frac14$, $W=2$ and $W=12$ (top to bottom).
  Shown is the pressure distribution scaled by its maximum value
  as a density plot over the lubrication zone.
  Only part of the domain is shown in the case with $\Xi=-15$.
  In (b) the pressures along the midline, $\zeta=0$,
  are plotted for the nine solutions, as well as
  for further solutions with
  ($\Xi_L=-\{1.5,2,3,4,5,7,15\}$, from blue to red).
  The dashed line in the top panel of (b) shows the
  narrow-wheel approximation, $-\frac18 \xi \eta^{-3} W^2$;
  those in the bottom panel indicate $\cP(\xi)$.
}
\end{figure}

Figure \ref{Qplot}(a) displays the
pressure distributions 
for three different wheel widths and bow-wave positions,
for steady planing solutions with $\eta_0=1$. When the gap
becomes relatively long ($|\xi_L|\gg1$), the pressure is localized to
the narrowest parts of the gap. In fact, the pressure distribution
becomes largely independent of $\xi_L$ once the bow wave
reaches a dimensionless distance of five or so from
the minimum gap. This feature is shown more clearly in
figure \ref{Qplot}(b), which displays pressure at wheel centre for
the same solutions as in figure \ref{Qplot}(a), together with additional ones
for different $\xi_L$.

When the wheel is relatively narrow, the pressure distribution
converges to a parabolic-in-$\zeta$ profile given by \eqref{4.5}.
As seen in the top panel of figure \ref{Qplot}(b), the central pressure aligns
well with this approximation for $W=\frac14$.
Wider wheels, on the other hand, lead to pressures that
are mostly uniform in $\zeta$, except in boundary layers at the
wheel edges ({\it cf.} the bottom row of panels in figure \ref{Qplot}(a)).
Indeed, throughout most of the lubrication zone,
$\cP \approx \Pi$ when $W\gg1$, as seen in the bottom
panel of figure \ref{Qplot}(b) for $W=12$.
%The boundary layers at the edges are easily detected in the
%series solution (\ref{3.6}), which reduces to
%\be
%\phi(\xi,\zeta) \approx - \sum_{j=1}^\infty c_j X_j(\xi)
%e^{-\lambda_j(\frac12 W - |\zeta|)}
%%\approx - c_1 X_1(\xi) e^{-\lambda_1(\frac12 W - |\zeta|)}
%.
%\ee
%Except near $\zeta=\pm\frac12W$, the sum is dominated by the first term
%provided the gap is not too long ($|\Xi_L|$ is not large).
% When the gap is long, the approximation
%$\phi\sim-\Pi e^{-(\frac12 W - |\zeta|)/C}$ is superior.

Given such solutions of the Reynolds equation, we may compute
the dimensionless load per unit width:
\begin{align}
\cL(\Xi_L,W) &= \frac{2}{W} \int_{\xi_L}^{\xi_R} \int_0^{W/2}
   % [\Pi(\xi) + \phi(\xi,\zeta)]
\cP(\xi,\zeta) \dd\zeta \dd\xi
%=  %\sum_{j=1}^\infty d_j =
%\cH^{-1}, 
\cr
%\be
% = 
%\int_{\Xi_L}^{\Xi_R} \Pi(\xi) \dd\xi
%- \frac{2}{W}  \sum_{j=1}^\infty c_j 
%\int_{\Xi_L}^{\Xi_R} X_j  \dd\xi \int_0^{W/2}
%  \frac{\cosh\lambda_j\zeta}{\cosh\half\lambda_jW} \dd\zeta
%\ee
%d_j
&=
\sum_{j=1}^\infty c_j \left( 1 - \frac{2\tanh\half\lambda_jW}{\lambda_jW} \right)
\int_{\xi_L}^{\xi_R} \phi_j  \dd\xi 
.
\label{Lforce}
\end{align}
The evolution equation for the bow wave also now boils down to
\be
\eta_{in} = \eta_L - \eta_L^3 [\Pi(\xi_L)'+\overline\phi_\xi(\xi_L)]
.
\label{fuxy}
\ee
Thus, for convenience,
we may fix $\eta_0=1$ and prescribe $\xi_L$ to compute the steady states,
rather than setting the load and flux and computing the corresponding
minimum gap and bow-wave
position, {\it via} \eqref{Lforce} and \eqref{fuxy}.

\begin{figure}%[htb!]
\begin{center}
\includegraphics[width=.85\linewidth]{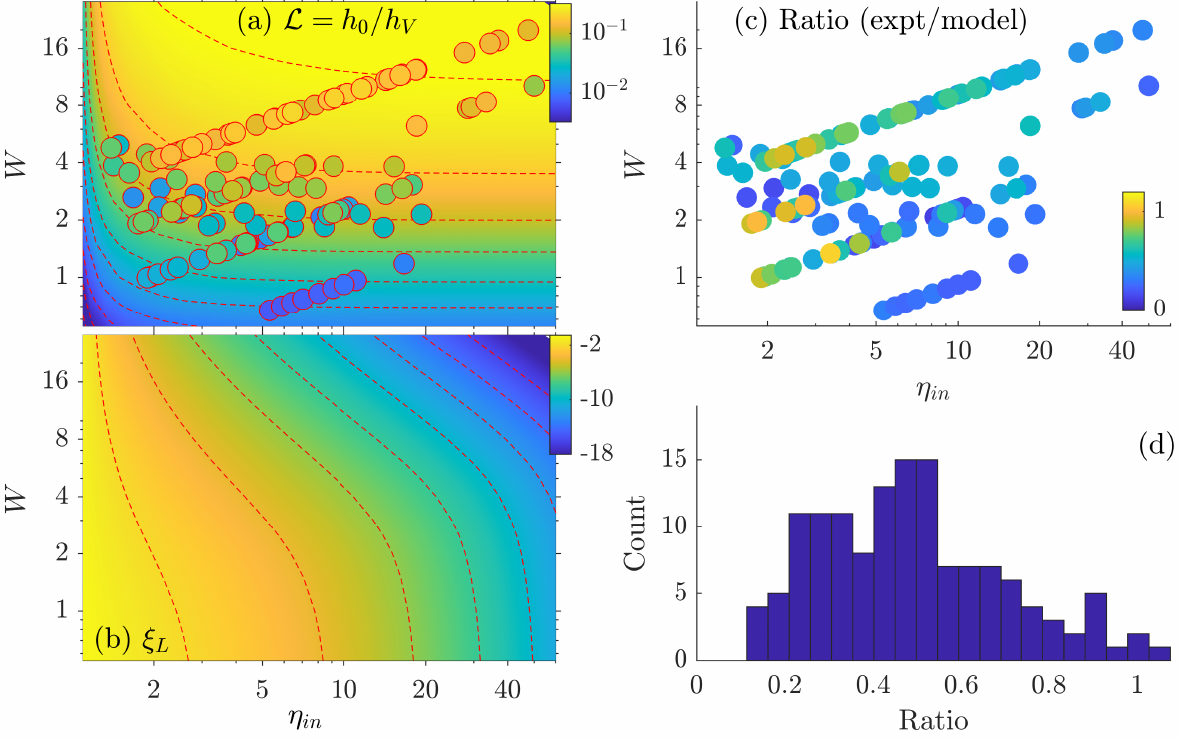}
\end{center}
\caption{\label{GGplo}
  Plots of (a) ${\cal L}$ and (b) $\xi_L$ as
  densities over the $(\eta_{in},W)-$plane.
  The circles indicate experimental measurements.
  The dashed lines in (a) show contours every quarter decade;
  those in (b) show those at negative even integers.
  In (c) and (d), the ratio of observed to predicted values of $h_0/h_V$
  is presented.
}
\end{figure}

Figure \ref{GGplo} presents numerical
results showing how the dimensionless load
and bow wave position vary over the $(\eta_{in},W)-$plane.
This figure also includes a set of experimental data,
generated from the laboratory experiment described in
\cite{rollpool}. For this data set, the fluid properties
and the wheel load and geometry
are all prescribed; the experiment is then run to find
the minimum gap. The results are then translated
into a dimensionless load and plotted on
the $(\eta_{in},W)-$plane. The degree to which the model matches the
experiment is shown further in figure \ref{GGplo}(c,d),
which shows the ratio of the theoretical prediction
for the load to that measured in the experiments.
Notably, the theory overpredicts the load by a factor
of two or so for most of the experiments.
This issue is also seen in figure \ref{sketch} which includes plots of
the minimum gap measured by a proximity sensor during three experiments
representative of relatively narrow, intermediate and wide wheels
($W=\{1.2,4.2,18.8\}$). To match the minimum gap over the steady
planing regime (as in panels (b-d)), the dimensionless load predicted
by the model is two to
three times larger than that imposed experimentally.
We return to this discrepancy and discuss it more thoroughly
below.

\subsection{Lift-off and touch-down}

\begin{figure}
\begin{center}
\includegraphics[width=.9\linewidth]{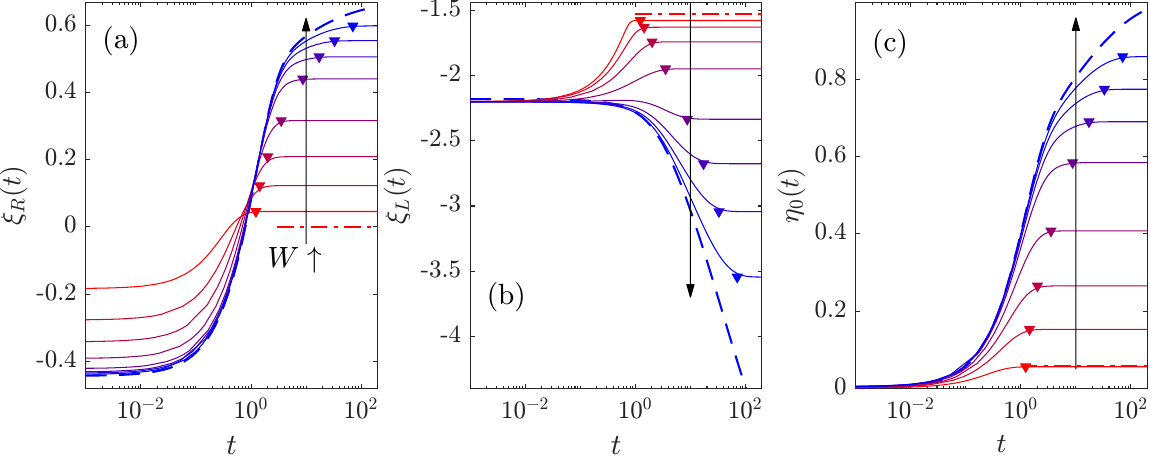}
\end{center}
\caption{\label{ivplot}
  Lift-off for solutions with varying, finite width, showing
  (a) $\xi_R(t)$, (b) $\xi_L(t)$ and (c) $\eta_0(t)$.
  The flux and load are fixed as those giving a minimum gap of unity and
  bow wave at $\xi_L=-5$ for an infinitely wide wheel
  ($\cL_0=0.339$ and $\eta_{in}=1.22$;
  {\it cf.} figure \ref{sold}); $\eta_0(0)=M=10^{-3}$, with $\xi_L(0)$
  given by \eqref{ics2}.
  The widths are
  $W=\{\frac16,\frac12,1,2,5,10,20,50\}$ (colour-coded from red to blue).
  The dashed lines show the solution for $\to\infty$, and the dot-dashed
  lines show the steady planing state predicted by \eqref{nss}.
  The triangles again show the time $t=3t_\infty$, with $t_\infty$ from \eqref{tinf}
  (using the final value of $\xi_L$ from each computation).
}
\end{figure}

Dynamical calculations of lift off for wheels with different width
are shown in figure \ref{ivplot}. In these examples, the
initial conditions are given by \eqref{ics2}, and the flux and load
are set equal to those that give a steady planing state
for an infinitely wide wheel with a minimum gap of unity and
bow wave at $\xi_L=-5$ ({\it cf.} figure \ref{sold}).
For the solution shown with the largest width, the minimum gap
and borders of the lubrication zone all follow a similar path
to those for an infinitely wide wheel, except that steady planing
with a shorter bow wave is reached sooner due to side flux.
As the wheel width narrows, the bow wave is no longer pushed
to the left, but instead migrates towards the minimum gap, and
steady planing is achieved with $\xi_R\approx0$ and
$\eta_0$ and $\xi_L$ approximately given by the narrow-wheel
results in \eqref{nss}. The time taken to reach the steady state
is again roughly estimated by $t=3t_\infty$, with $t_\infty$ from \eqref{tinf}
(see the times marked by triangles in figure \ref{ivplot}).

\begin{figure}
\begin{center}
\includegraphics[width=.9\linewidth]{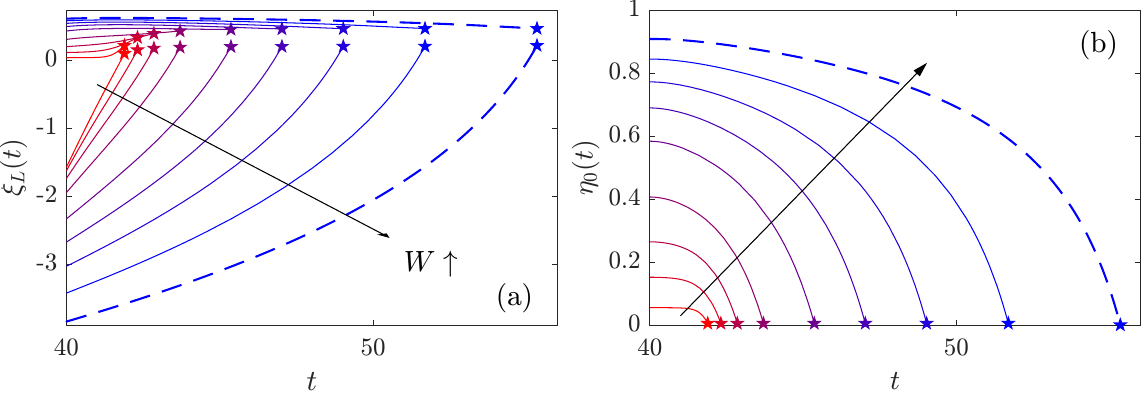}
\end{center}
\caption{\label{ivplot2}
  Touch-down for the solutions show in figure \ref{ivplot}, after the
  flux is switched off at $t=40$.
}
\end{figure}

When the flux is switched off at $t=40$, the solutions from figure
\ref{ivplot} descend to touch down, as shown in figure \ref{ivplot2}.
Again, touch-down arises at contact with a finite lubrication
zone to the right.
For the narrower wheels, the touch-down time is close to the
time taken to translate the bow wave to the minimum gap
($t-t_e\sim |\Xi_L(t_e)|$), as noted in \S\ref{narrow};
this time becomes more prolonged as $W$ increases.
Overall, as intended, the finite-width model bridges between the
limits of a narrow and a wide wheel.

Figure \ref{sketch}(b-d) displays how the predictions of 
the model for lift-off, steady planing and touch-down compare with
measurements from three experiments, one relatively narrow
(but not particularly so), one
of intermediate width, and one relatively wide.
The three phases of evolution are qualitatively captured. However,
in addition to the mismatch between the dimensionless load values,
the duration of lift-off and touch-down are much longer for the narrow
wheel in the experiments, whereas touch-down takes much longer in the
model for the wide wheel. These disrepancies in the dynamics suggest that
the development and length of the bow wave may not be captured
adequately in the model.

\section{Discussion}\label{sec:disc}

In this paper, we derived a mathematical model to describe the dynamics of levitation of a wheel rolling over a pool of viscous fluid.
The model was first solved in two asymptotic limits. For an infinite wide wheel,
the model reduces to a system of non-linear differential equations in time and one space dimension (the rolling direction). Both steady planing and back-flowing
flooding solutions are possible, depending on the level of the incoming flux.
We further examined the process of lift-off, the approach to steady planing
or flooding, and the final touch-down.
For a narrow wheel, the model again simplifies to
a system of non-linear differential equations. In this limit,
only steady planing states are possible,
the flooding solution becoming eliminated by side flux. 
Again, we explored lift-off and touch-down, finding
similar dynamics to that seen for an infinitely wide wheel.

To bridge the gap between these two limits, we investigated wheels of
finite width after introducing some additional approximations. More specifically,
to avoid the complicated free boundary problem, we took a wheel average of the
boundary conditions at the front and back and then adopted a rectangular domain.
%Further, an iterative scheme was developed to help locate the optimal value of the front boundary.
The Reynolds lubrication equation could then be solved in a straightforward
manner using a truncated series solution derived from separation of variables.
This model reduces to the cases of wide or narrow wheels in suitable limits,
and predicts characteristics for steady planing, lift-off and touch-down.
Note that both lift-off and touch down take place in a continuous
fashion in the model because the lubrication zone migrates to localized
regions lying to the right of the minimum gap. The
lubrication pressure then always remains finite because the gap
never closes there. This avoids any issues associated with
diverging squeeze-flow forces over closing gaps \cite{brenner,contactI}.

It is possible to proceed further with the full model in order to
gauge the fidelity of 
the approximation in which the lubrication zone is treated as
rectangular for finite wheel width. This task was taken up in our Appendix.
There, we showed how one could perturbatively build the shape of the
film-splitting position for steady planing solutions. This exercise
shows how the deviation of that shape from a straight line is
relatively mild, suggesting that the approximate treatment of the right-hand
edge of the lubrication zone may be adequate. By contrast,
the bow wave to the left of the lubrication zone likely has
a rather more strongly distorted shape, suggesting that the
approximation there is less trustworthy.

We also explored how well the wheel-averaged model of \cite{rollpool}
reproduced our results for wheels of finite width. Because the
same rectangular approximation of the geometry of the lubrication zone is adopted
during wheel averaging, a demanding test of that model is that it
reproduce our finite-width results. In order for this to be true,
however, we found that the free parameter present in the wheel-averaged
model should actually depend on wheel width and bow-wave position.
Worse, our calibrations of this parameter turn out to be very different
from the values found from fitting the parameter to experimental
data \cite{rollpool}. Evidently, the wheel-averaged model
cannot be fully trusted. One explanation for the apparent
success of the fitted model in \cite{rollpool} is
that the fitted value somehow accounted for some omitted
relevant physical effects. That said, the data reported by \cite{rollpool}
appear to have been affected by a problem with the bearings
of the system used to hold the wheel in place under the
applied load. More recent experiments, in which this bearing problem
was fixed, suggest that the fits may not be reliable.

Finally, we compared the results of the model with steady-planing data
extracted from experiments using the set-up of \cite{rollpool}
(having fixed the bearings).
We found significant discrepancies between theory and observation, adopting
the minimum gap measured and other experimental variables
to set the model parameters then comparing the load predicted with
that prescribed experimentally.
Overall, there is roughly a factor of two between the measured and
predicted loads. 
Time series of the minimum gap measured by a proximity sensor
during the experiments also indicate that the model can
fail to adequately predict the duration of lift-off and touch-down.
Such discrepancies point to a flaw in the model.
The most obvious limitation seems to be the treatment of
the bow wave and assuming that it is largely planar.
Indeed, videos taken during the experiments
also suggest that the bow wave possesses significant shape across the wheel. 

\appendix

\section{Further details of the steady planing states with finite wheel width}

\subsection{Evaluation of the left boundary condition}\label{BPL}

At the left boundary, with the full form of the flux condition in
(\ref{2.5}), we set $\xi_L=\Xi_L+\Delta_L(\zeta)$, where $\Xi_L$ denotes
the constant position predicted by \eqref{fuxy} and $\Delta_L$
represents any spatial variation of the bow wave. 
The full flux condition now translates to
\be
\sqrt{1+\half(\Delta_L')^2} (\eta(\xi_L) - \eta_{in}) = \eta(\xi_L)^3
\left.\left(\Pi_\xi + \phi_\xi - \Delta_L' \phi_\zeta\right)
\right|_{\xi=\xi_L}
.\label{3.10}
\ee
To leading order, this would predict that
\be
\eta_{in} \approx N_{in} =
\eta_L - \eta_L^3 \Pi'(\Xi_L)  - \eta_L^3 \phi_\xi(\Xi_L,\zeta),
\label{2.11}
\ee
where $\eta_L=\eta(\Xi_L)$. 

\begin{figure}
\begin{center}
\includegraphics[width=.8\linewidth]{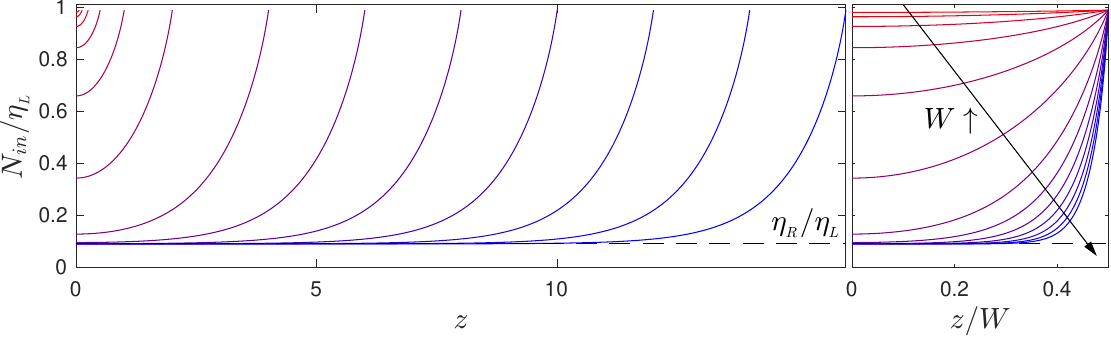}
\end{center}
\caption{\label{BPL0}
  Numerical results showing the function $N_{in}(\zeta)$
  defined in (\ref{2.11})
  for $(\Xi_L,\Xi_R)=(-5,0.662)$ 
  and various wheel widths
  ($W=\frac18,\frac14,\frac12,1,2,4,8,12,20,24,28,32$, from red to blue).
}
\end{figure}

To gauge the fidelity of the approximation in which we
omit $\Delta_L$, we examine
the function $N_{in}(\zeta)$ from \eqref{2.11} in more detail. Sample
computations of this function are shown in figure \ref{BPL0}
for $(\Xi_L,\Xi_R)=(-5,0.662)$ and varying wheel width.
For narrower wheels, $N_{in}$ is close to a constant, confirming that
the approximation works well. This result is evident from
the results in \S\ref{narrow}, which indicate
that $\cP=O(W^2)$ in this limit, and so $N_{in}\approx\eta_L$
(as seen in figure \ref{BPL0}).

With a relatively wide wheel, on the other hand, the boundary-layered
structure to the pressure distribution implies that
$\cP\approx\Pi$ and $\phi=O(W^{-1})$
over most of the lubrication zone; only in
the boundary layers at the wheel's edges does $\cP$ fall to zero
and $\phi$ become order one (figure \ref{Qplot}; bottom row).
Consequently, $N_{in}\approx \eta_L-\eta_L^3\Pi' = \eta_R$,
except near the wheel's edges, where the function
necessarily increases up to $N_{in}=\eta_L$.
The approximation therefore breaks down near the left-hand corners
of the lubrication zone, which presumably become
rounded off. Setting aside those boundary layers, however,
one concludes that the position of the bow wave is again
independent of $\zeta$.

In figure \ref{BPL0},
the approximation is arguably worst for $W=4$: for this
case, there is an almost
one hundred percent variation of $N_{in}$ about its
mean value, with only a narrow central region over which the
function is nearly constant.
Thus, except in the narrow and wide limits, the assumption
that the bow wave is straight is suspicious.

\subsection{Perturbing the right boundary}\label{sec:RBP}

\begin{figure}
\begin{center}
\includegraphics[width=.7\linewidth]{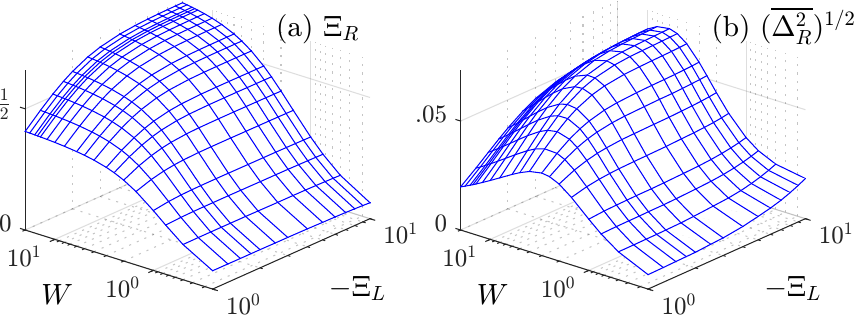}
\end{center}
\caption{\label{BPRit}
  (a) The position of the right-hand border of the lubrication
  region $\Xi_R$ as a function of wheel with $W$ and left-hand
  border position $\Xi_L$. The root-mean-square average value
  of the residual, $(\overline{\Delta_R^2})^{1/2}$, is shown in
  (b), and in panel (c) we compare the ratio of $\Xi_R$ with its
  un-iterated value $\Xi_{R*}$. The dashed lines indicate the value of
  $\Xi_R$ used to generate the edge positions.
}
\end{figure}

\begin{figure}
\begin{center}
\includegraphics[width=.8\linewidth]{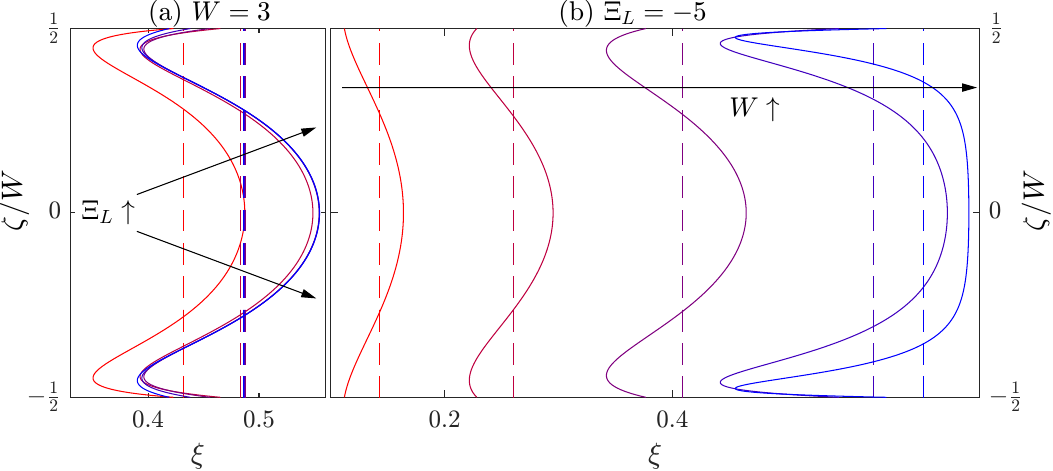}
\end{center}
\caption{\label{BPRit2}
  Predicted shapes for the right edge of the lubrication
  region for (a) $W=3$ and $\xi_L=-\{\frac32,3,5,7.5,10\}$,
  and (b) $\Xi_L=-5$ and $W=\{\frac12,1,2,6,12\}$
  (in each case, colour-coded from red to blue).
  The dashed lines show $\xi=\Xi_R$.
}
\end{figure}

Our appromixation at the right-hand edge of the lubrication zone
sets $\xi_R=\Xi_R$ as the position where $\overline\cP_\xi=0$.
In further detail, we may set $\xi_R=\Xi_R+\Delta_R(\zeta)$,
where $\Delta_R(\zeta)$ denotes the spatially varying piece of border
when we return to the full boundary condition
$\cP_\xi(\xi_R,\zeta)=0$.  Assuming that $|\Delta_R|\ll1$,
that boundary condition can be Taylor expanded to find
\be
\Delta_R = - \left. \frac{\cP_\xi}{\cP_{\xi\xi}}\right|_{\xi=\Xi_R}.
\label{3.17}
\ee
Figures \ref{BPRit} and \ref{BPRit2} display computations of $\Xi_R$
and $\Delta_R$ for a range of wheel widths $W$
and bow-wave positions $\Xi_L$. The average residual,
$(\overline{\Delta_R^2})^{1/2}$, is less than
twenty percent of $\Xi_R$, except for the smallest widths
and largest values of $|\Xi_R|$. Hence, it appears that it is a
fair approximation to treat the right-hand
border of the lubrication zone as straight.

\begin{figure}
\begin{center}
\includegraphics[width=.75\linewidth]{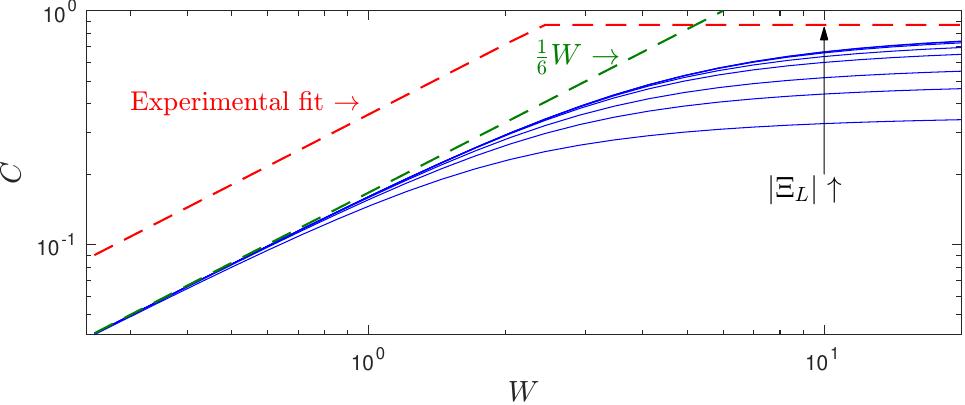}
\end{center}
\caption{\label{Citplot}
  The constant $C$ defined in (\ref{3.22}) as a function of $W$ for 
  $\Xi_L = -\{10,8,6,4,3,2,\frac32\}$.
  The dashed lines show the narrow-wheel limit $C\sim\frac16W$
  and the experimental fit reported in \cite{rollpool}.
}
\end{figure}

\subsection{Assessing a wheel-averaged model}
\label{reassess}
In \cite{rollpool}, the full lubrication problem was approximated
by a simple ordinary differential equation in $\xi$ by averaging
over the wheel width. This approximation relied on a further assumption
that the side flux at the wheel edges, or equivalently the
transverse pressure gradient $\cP_\zeta$, could be
approximated using the wheel-average pressure $\overline{\cP}$.
For our series solution,
\be
\cP_\zeta(\xi,\half W)
= - \sum_{j=1}^\infty c_j \lambda_j \phi_j(\xi) \tanh \left(\half\lambda_j W\right)
\label{3.19}
\ee
and
\be
\overline{\cP}(\xi) = \Pi(\xi)
- \frac{2}{W} \sum_{j=1}^\infty c_j \lambda_j^{-1} \phi_j(\xi)
\tanh \left(\half\lambda_j W\right)
.
\label{3.20}
\ee
The approximation is to set
\be
\overline\cP(\xi) = - C \cP_\zeta(\xi,\half W)
,
\ee
where $C$ is a wheel-width dependent constant.

We evaluate this approximation by computing $C$ using the
least-squares estimate,
\be
C = -
\frac{
\int_{\Xi_L}^{\Xi_R} \cP_\zeta(\xi,\half W) \overline{\cP}(\xi) \;\dd\xi
}{
  %\left\{
  \int_{\Xi_L}^{\Xi_R} [\cP_\zeta(\xi,\half W)]^2 \;\dd\xi
%\right\}^{-1}
}\label{3.22}
.
\ee
This choice actually leads to a constant $C$ that depends on
both $W$ and $\Xi_L$, as illustrated in figure \ref{Citplot}.
However, once the left edge becomes sufficiently distant
($\Xi_L<-3$), the dependence on that parameter is relatively weak.
The approximation works well for $W\ll1$, with $C\sim \frac16W$,
as expected from \S\ref{narrow}.

The values of $C$
predicted theoretically are not those that are inferred when one
attempts to match theoretical predictions of the minimum gap or
load with those observed experimentally, treating $C$ as a fitting parameter
\cite{rollpool}. The fits indicated that $C\approx0.87$ for $W>2.5$
and $C\approx0.36W$ for $W<2.5$, and are included in
figure \ref{Citplot}. The latter differs from
the narrow-wheel asymptotic limit $C\sim \frac16W$
by a factor of about two, whereas the former seems inconsistent
with the theoretical predictions for wider wheels,
given the range of the experimental data
indicated by figure \ref{GGplo}.

%Note that the wheel-averaged approximation is also equivalent to
%setting
%\be
%\cP \sim \left[ 1 - \frac{\cosh \zeta/C }{\cosh W/2C} \right]
%\frac{\overline{\cP}}{1-\tanh W/2C},
%\ee
%???

\bibliographystyle{unsrt}
\bibliography{hat}

\end{document}